\documentclass[prd,aps,showpacs,epsf,floats,onecolumn]{revtex4}%
\usepackage{amssymb}
\usepackage{amsfonts}
\usepackage{amsmath}
\usepackage{graphicx}%
\providecommand{\U}[1]{\protect\rule{.1in}{.1in}}

\begin{document}
\title{\textbf{Information Geometric Complexity of Entropic Motion on Curved
Statistical Manifolds under Different Metrizations of Probability Spaces}}
\author{\textbf{Steven Gassner} and \textbf{Carlo Cafaro}}
\affiliation{SUNY Polytechnic Institute, 12203 Albany, New York, USA}

\begin{abstract}
We investigate the effect of different metrizations of probability spaces on
the information geometric complexity of entropic motion on curved statistical
manifolds. Specifically, we provide a comparative analysis based upon
Riemannian geometric properties and entropic dynamical features of a Gaussian
probability space where the two distinct dissimilarity measures between
probability distributions are the Fisher-Rao information metric and the
$\alpha$-order entropy metric. In the former case, we observe an asymptotic
linear temporal growth of the information geometric entropy (IGE) together
with a fast convergence to the final state of the system. In the latter case,
instead, we note an asymptotic logarithmic temporal growth of the
IGE\ together with a slow convergence to the final state of the system.
Finally, motivated by our findings, we provide some insights on a tradeoff
between complexity and speed of convergence to the final state in our
information geometric approach to problems of entropic inference.

\end{abstract}

\pacs{Chaos (05.45.-a), Complexity (89.70.Eg), Entropy (89.70.Cf), Inference Methods
(02.50.Tt), Information Theory (89.70.+c), Probability Theory (02.50.Cw),
Riemannian Geometry (02.40.Ky).}
\maketitle

\bigskip\pagebreak

\section{Introduction}

Methods of information geometry \cite{amari,amari2,amari3} can be combined
with entropic inference techniques \cite{caticha12} to quantify the complexity
of statistical models used to render probabilistic descriptions of systems
about which only limited information is known. Within this hybrid framework,
the complexity associated with statistical models can be viewed as a measure
of the difficulty of inferring macroscopic predictions due to the lack of
complete knowledge about the microscopic degrees of freedom of the system
being analyzed \cite{cafarophd}. Initially, entropic methods can be employed
to establish an initial, static statistical model of the complex system. Then,
after identifying the microscopic degrees of freedom of a complex system and
selecting its relevant information constraints, the statistical model that
characterizes the complex system is specified by means of probability
distributions parametrized in terms of statistical macrovariables. These
variables, in turn, depend upon the specific functional expression of the
information constraints assumed to be important for implementing statistical
inferences about the system of interest. Once the probability space is endowed
with a suitable notion of metric needed to distinguish different elements of
the statistical model, one focuses on the evolution of the complex system.
Specifically, assuming the complex system evolves, the evolution of the
associated statistical model from its initial to final configurations can be
determined by means of the so-called Entropic Dynamics (ED, \cite{catichaED}).

Entropic Dynamics is a form of information-constrained dynamics on curved
statistical manifolds whose elements are probability distributions. Moreover,
these distributions are in one-to-one relation with a convenient set of
statistical macrovariables that specify a parameter space which provides a
parametrization of points on the statistical manifold. The ED setting
specifies the evolution of probability distributions in terms of an entropic
inference principle:\textbf{\ }starting from the initial configuration, the
motion toward the final configuration occurs via the maximization of the
logarithmic relative entropy functional (Maximum relative Entropy method- MrE
method, \cite{caticha12,adom06,adom07,adomphd}) between any two consecutive
intermediate configurations of the system. ED\ generates only the
\emph{expected}, but not the \emph{actual}, trajectories of the system. In
this regard, we stress that uncovering links between information geometry and
classical Newtonian mechanics can be of great theoretical interest
\cite{caticha07,nico18}. For instance, a formal bridge between information
geometric techniques and classical dynamical systems was recently proposed by
using the concept of canonical divergence in dually flat manifolds in Ref.
\cite{nico18}. Inferences within ED rely on the nature of the selected
information constraints that are employed at the level of\textbf{\ }the MrE
algorithm. Modeling schemes of this type can only be validated \emph{a
posteriori}. If discrepancies occur between inferred predictions and
experimental observations, a new set of information constraints must be chosen
\cite{jaynes85,dewar09,giffin16}. This is an especially important feature of
the MrE algorithm and was recently reconsidered in Ref. \cite{cafaropre16} by
applying entropic inference techniques to stochastic Ising models. The above
mentioned entropic maximization procedure specifies the evolution of
probability distributions as a geodesic evolution of the statistical
macrovariables \cite{caticha12}. For recent reviews on an information
geometric perspective on the complexity of macroscopic predictions arising
from incomplete information, we refer to Refs. \cite{ali17,felice18,ali18}.

A common measure of distance between two different probability distributions
is quantified by the Fisher-Rao information metric \cite{amari}. This distance
can be regarded as the degree of distinguishability between two probability
distributions. After having determined the information metric, one can apply
the usual methods of Riemannian differential geometry to study the geometric
structure of the statistical manifold underlying the entropic motion which
determines the evolution of the probability distributions. Conventional
Riemannian geometric quantities such as Christoffel connection coefficients,
the Ricci tensor, the Riemannian curvature tensor, sectional curvatures,
scalar curvature, the Weyl anisotropy tensor, Killing fields, and Jacobi
fields can be computed in the usual manner \cite{thorne73}. Furthermore, the
chaoticity (that is, temporal complexity) of such statistical models can be
analyzed in terms of convenient indicators, such as: the signs of scalar and
sectional curvatures of the statistical manifold, the asymptotic temporal
behavior of Jacobi fields, the existence of Killing vectors, and the existence
of a non-vanishing Weyl anisotropy tensor. In addition to these measures,
complexity can also be quantified\textbf{\ }by means of the so-called
information geometric entropy (IGE, \cite{ali17,felice18,ali18}).

From a theoretical standpoint, the utility of the Fisher-Rao information
metric as a suitable distinguishability measure of two probability
distribution functions is mainly motivated by Cencov's theorem
\cite{cencov,campbell}. This theorem states that the Fisher-Rao information
metric is, modulo an unimportant constant factor, the only Riemannian metric
that is invariant under mappings referred to as congruent embeddings by Markov
morphisms \cite{caticha12}. From a computational standpoint, however, the
algebraic form of the Fisher-Rao information metric makes it rather difficult
to use when applied to multi-parameter spaces like mixture models. For
instance, a fundamental drawback of the Fisher-Rao metric is that it is not
available in closed-form for a mixture of Gaussians \cite{peter06}. These
computational inefficiencies extend to the computation of the Christoffel
connection coefficients and, therefore, to the integration of geodesic
equations. The challenges with the mixture models were originally encountered
in the framework of shape matching analysis of medical and biological image
structures where shapes are represented by a mixture of probability density
functions \cite{peter06}. To partially address the above mentioned
computational issues, a different Riemannian metric based upon the generalized
notion of $\phi_{\alpha}$-entropy functional was employed
\cite{havrda67,burbea82}. The corresponding $\alpha$-order entropy metric
allows us to obtain closed-form solutions to both the metric tensor and its
derivatives for the Gaussian mixture model. Thus, compared to the Fisher-Rao
information metric, the $\alpha$-order entropy metric enhances the
computational efficiency in shape analysis tasks \cite{peter06}.

In this paper, inspired by the above-mentioned enhanced computational
efficiency of the $\alpha$-order entropy metric with respect to the Fisher-Rao
information metric, we seek to address the following questions: i) How does a
different choice of metrization of probability spaces affect the complexity of
entropic motion on a given probability space? ii) Does a possible higher
computational efficiency of the $\alpha$-order metric with respect to the
Fisher-Rao information metric lead to a lower information geometric complexity
of entropic motion? iii) Is there a tradeoff between speed of convergence to
the final macrostate and the information geometric complexity of entropic
motion? \ 

Our motivation to explicitly compute geometrical quantities including the
scalar curvature, the sectional curvature, the Ricci curvature tensor, the
Riemann curvature tensor, and the Weyl anisotropy tensor is twofold. First, we
wish to present here a comparative analysis of both geometrical and entropic
dynamical nature between the Fisher-Rao and the $\alpha$-metrics.\ Second, in
view of possible further investigations concerning the geodesic deviation
behavior on curved statistical manifolds along the lines of those presented in
Refs. \cite{physicad1,physicad2}, having the explicit expressions of such
geometrical quantities can be quite useful for future efforts. In this
respect, for instance, our result concerning the maximal symmetry of the
Gaussian probability space endowed with the $\alpha$-metric can have important
implications when integrating the Jacobi geodesic spread equation in order to
study the deviation of two neighboring geodesics on the manifold
\cite{casetti}. Indeed, for maximally symmetric manifolds, the sectional
curvature (that is, the generalization to higher-dimensional manifolds of the
usual Gaussian curvature of two-dimensional surfaces) assumes a constant value
throughout the manifold. As a result, exploiting this symmetry reduces
significantly the otherwise challenging problem of integrating the Jacobi
deviation equation by simplifying the differential equation via the expression
of the Riemann curvature tensor components that enter it.

The layout of the remainder of the paper is as follows. In Section II, we
briefly present the Fisher-Rao information and the $\alpha$-order entropy
metrics as special cases of the so-called $\phi_{\alpha}$-entropy metric. In
Section III, we describe the information geometric properties of Gaussian
probability spaces equipped with the above-mentioned metrizations. In Section
IV, we study the geodesics of the entropic motion on the two curved
statistical manifolds. In Section V, we report the asymptotic temporal
behavior of the information geometric entropy of both statistical models. Our
final remarks appear in Section VI. Finally, technical details can be found in
Appendix A.

\section{Metrizations of probability spaces}

In this section, we focus on two different metrizations of a probability
space. Specifically, we consider the Fisher-Rao information metric and the
$\alpha$-order entropy metric. These two metrics are limiting cases of a large
class of generalized metrics introduced by Burbea and Rao in Ref.
\cite{burbea82}.

For a formal mathematical discussion on the $\phi$-entropy functional
formalism, we refer to Ref. \cite{burbea82}. In what follows, we present a
minimal amount of information concerning this topic needed to follow our work.
A $\phi$-entropy metric $g_{ij}^{\left(  \phi\right)  }\left(  \theta\right)
$\ is formally defined as the Hessian of a $\phi$-entropy functional along a
direction of the tangent space of the parameter space $\mathcal{D}_{\theta}%
$\textbf{. }Note that $\theta\overset{\text{def}}{=}\left\{  \theta
^{k}\right\}  $\ with $1\leq k\leq N$\ and $N$\ being the dimensionality of
the parameter space. Specifically, we have%
\begin{equation}
g_{ij}^{\left(  \phi\right)  }\left(  \theta\right)  \overset{\text{def}}%
{=}-\frac{\Delta_{\theta}\left[  H_{\phi}\left(  p\right)  \right]  }%
{\partial\theta^{i}\partial\theta^{j}}=\int_{\mathcal{X}}\phi^{\prime\prime
}\left(  p\right)  \left(  \frac{\partial p}{\partial\theta^{i}}\right)
\left(  \frac{\partial p}{\partial\theta^{j}}\right)  dx\text{,} \label{la}%
\end{equation}
with $i$, $j=1$,..., $N$, and
\begin{equation}
H_{\phi}\left(  p\right)  \overset{\text{def}}{=}-\int_{\mathcal{X}}%
\phi\left(  p\right)  dx\text{.} \label{functional}%
\end{equation}
The quantity $H_{\phi}\left(  p\right)  $\ denotes a $\phi$-entropy
functional, $\Delta_{\theta}$ denotes the Hessian of $H_{\phi}\left(
p\right)  $\ along the direction $dp\overset{\text{def}}{=}\left(  \partial
p/\partial\theta^{k}\right)  d\theta^{k}$ where repeated indices are summed
over, $\phi\left(  p\right)  $\ is a generalized convex real-valued $C^{2}%
$-function, $p=p\left(  x|\theta\right)  $\ is a probability density function,
and\textbf{ }$\mathcal{X}$\textbf{\ }is the microspace of the system. The
quantity $\phi^{\prime\prime}\left(  p\right)  $\ in\ Eq. (\ref{la}) can be
formally regarded as the second derivative of the function $\phi$\ with
respect to $p$\ viewed as an ordinary real-valued variable. In
particular\textbf{, }when $\phi\left(  p\right)  $ is defined as%
\begin{equation}
\phi_{\alpha}\left(  p\right)  \overset{\text{def}}{=}\left\{
\begin{array}
[c]{c}%
p\log\left(  p\right)  \text{, if }\alpha=1\\
\\
\left(  \alpha-1\right)  ^{-1}\left(  p^{\alpha}-p\right)  \text{, if }%
\alpha\neq1
\end{array}
\right.  \text{,}%
\end{equation}
we obtain $\phi_{1}^{\prime\prime}\left(  p\right)  =1/p$ and $\phi
_{2}^{\prime\prime}\left(  p\right)  =2$ where the characteristic parameter
$\alpha$ equals $1$ and $2$, respectively. In the former case, $g_{ij}%
^{\left(  \phi_{1}\right)  }\left(  \theta\right)  $ reduces to the Fisher-Rao
information metric $g_{ij}^{\left(  \text{FR}\right)  }\left(  \theta\right)
$,%
\begin{equation}
g_{ij}^{\left(  \text{FR}\right)  }\left(  \theta\right)  \overset{\text{def}%
}{=}-\left(  \frac{\partial^{2}\mathcal{S}\left(  \theta^{\prime}\text{,
}\theta\right)  }{\partial\theta^{\prime i}\partial\theta^{\prime j}}\right)
_{\theta^{\prime}=\theta}=%
{\displaystyle\int\limits_{\mathcal{X}}}
\frac{1}{p\left(  x|\theta\right)  }\frac{\partial p\left(  x|\theta\right)
}{\partial\theta^{i}}\frac{\partial p\left(  x|\theta\right)  }{\partial
\theta^{j}}dx\text{.} \label{FR}%
\end{equation}
The quantity\textbf{ }$\mathcal{S}\left(  \theta^{\prime}\text{, }%
\theta\right)  $ in\ Eq. (\ref{FR}) denotes the relative entropy functional
given by \cite{caticha12},%
\begin{equation}
\mathcal{S}\left(  \theta^{\prime}\text{, }\theta\right)  \overset{\text{def}%
}{=}-\int_{\mathcal{X}}p\left(  x|\theta^{\prime}\right)  \log\left[
\frac{p\left(  x|\theta^{\prime}\right)  }{p\left(  x|\theta\right)  }\right]
dx\text{.}%
\end{equation}
In the latter case, $\frac{1}{2}g_{ij}^{\left(  \phi_{2}\right)  }\left(
\theta\right)  $ becomes the $\alpha$-order metric tensor with $\alpha=2$
given by%
\begin{equation}
g_{ij}^{\left(  \alpha\right)  }\left(  \theta\right)  \overset{\text{def}}{=}%
{\displaystyle\int\limits_{\mathcal{X}}}
\frac{\partial p\left(  x|\theta\right)  }{\partial\theta^{i}}\frac{\partial
p\left(  x|\theta\right)  }{\partial\theta^{j}}dx\text{.} \label{alpha}%
\end{equation}
In the next section, we employ the metrics in\ Eqs. (\ref{FR}) and
(\ref{alpha}) to measure the distance between probability distributions of a
Gaussian statistical manifold.

\section{Information geometry of a Gaussian statistical model}

In this section, we study the information geometry of a two-dimensional
probability space specified by Gaussian probability distributions. In the
first case, we assume the metrization is defined by the Fisher-Rao information
metric in Eq. (\ref{FR}). In the second case, instead, we assume the
metrization is given by the $\alpha$-order metric in Eq. (\ref{alpha}).

\subsection{The Fisher-Rao information metric}

Consider a single-variable Gaussian probability density function
$p(x\,|\,\mu_{x}$, $\sigma_{x})$ given by,%
\begin{equation}
p(x\,|\,\mu_{x}\text{, }\sigma_{x})=\frac{1}{\sqrt{2\pi\sigma_{x}^{2}}%
}e^{-\frac{(x-\mu_{x})^{2}}{2\sigma_{x}^{2}}}\text{.} \label{steven1}%
\end{equation}
For a Gaussian distribution, we let $\theta=(\theta^{1}$, $\theta^{2}%
)=(\mu_{x}$, $\sigma_{x})$ with $\mu_{x}\in%
\mathbb{R}
$ and $\sigma_{x}\in%
\mathbb{R}
_{+}\backslash\left\{  0\right\}  $. Therefore, the two-dimensional Gaussian
statistical manifold $\left(  \mathcal{M}_{s}\text{, }g\right)  $ is such
that,%
\begin{equation}
\mathcal{M}_{s}\overset{\text{def}}{=}\left\{  p(x\,|\,\mu_{x}\text{, }%
\sigma_{x})\text{ in\ Eq. (\ref{steven1})}:\mu_{x}\in%
\mathbb{R}
\text{ and }\sigma_{x}\in%
\mathbb{R}
_{+}\backslash\left\{  0\right\}  \right\}  \text{,}%
\end{equation}
with $g$ being the selected metric. Substituting Eq. (\ref{steven1}) into Eq.
(\ref{FR}), we obtain
\begin{equation}
g_{ij}^{(\text{FR})}\left(  \mu_{x}\text{,}\sigma_{x}\right)  =\frac{1}%
{\sigma_{x}^{2}}%
\begin{pmatrix}
1 & 0\\
0 & 2
\end{pmatrix}
\text{.} \label{steven3}%
\end{equation}
Using the metric tensor components in Eq. (\ref{steven3}), we can study a
variety of global properties of the two-dimensional Gaussian statistical
manifold. For instance, the affine connection coefficients (also known as the
Christoffel symbols of second kind) are defined as
\cite{defelice90,weinberg72},%
\begin{equation}
\Gamma_{ij}^{k}\overset{\text{def}}{=}\frac{1}{2}g^{km}(\partial_{i}%
\,g_{mj}+\partial_{j}\,g_{im}-\partial_{m}\,g_{ij})\text{.} \label{steven4}%
\end{equation}
The quantity $g^{ij}$ in Eq. (\ref{steven4}) is such that $g^{ij}g_{jk}%
=\delta_{k}^{i}$ where $\delta$ denotes the Kronecker delta,%
\begin{equation}
\left(  g^{(\text{FR})}\right)  ^{ij}\left(  \mu_{x}\text{,}\sigma_{x}\right)
=\sigma_{x}^{2}%
\begin{pmatrix}
1 & 0\\
0 & \frac{1}{2}%
\end{pmatrix}
\text{.} \label{steven5}%
\end{equation}
Substituting Eqs. (\ref{steven3}) and (\ref{steven5}) into Eq. (\ref{steven4}%
), we get%
\begin{equation}
\Gamma_{11}^{1}=0\text{, }\Gamma_{12}^{1}=\Gamma_{21}^{1}=-\frac{1}{\sigma
_{x}}\text{, }\Gamma_{22}^{1}=0\text{, }\Gamma_{11}^{2}=\frac{1}{2\sigma_{x}%
}\text{, }\Gamma_{22}^{2}=-\frac{1}{\sigma_{x}}\text{, and }\Gamma_{12}%
^{2}=\Gamma_{21}^{2}=0\text{.} \label{steven6}%
\end{equation}
These connection coefficients in Eq. (\ref{steven6}) allow us to quantify the
curvature properties of the statistical manifold. Let us first consider the
Riemann curvature tensor $\mathcal{R}_{ijk}^{l}$ \cite{defelice90,weinberg72}%
,
\begin{equation}
\mathcal{R}_{ijk}^{l}\overset{\text{def}}{=}\partial_{j}\,\Gamma_{ki}%
^{l}-\partial_{k}\,\Gamma_{ji}^{l}+\Gamma_{jm}^{l}\Gamma_{ki}^{m}-\Gamma
_{km}^{l}\Gamma_{ji}^{m}\text{.} \label{steven7}%
\end{equation}
Substituting Eq. (\ref{steven6}) into Eq. (\ref{steven7}), the non-vanishing
Riemann curvature tensor components are%
\begin{equation}
\mathcal{R}_{212}^{1}=-\frac{1}{\sigma_{x}^{2}}\text{, }\mathcal{R}_{221}%
^{1}=\frac{1}{\sigma_{x}^{2}}\text{, }\mathcal{R}_{112}^{2}=\frac{1}%
{2\sigma_{x}^{2}}\text{, and }\mathcal{R}_{121}^{2}=-\frac{1}{2\sigma_{x}^{2}%
}\text{.} \label{ciao1}%
\end{equation}
Furthermore, the Ricci curvature tensor $\mathcal{R}_{ij}$ is given by,%
\begin{equation}
\mathcal{R}_{ij}\overset{\text{def}}{=}\mathcal{R}_{ikj}^{k}=\partial
_{k}\,\Gamma_{ij}^{k}-\partial_{j}\,\Gamma_{ik}^{k}+\Gamma_{ij}^{k}\Gamma
_{kn}^{n}-\Gamma_{ik}^{m}\Gamma_{jm}^{k}\text{.} \label{steven8}%
\end{equation}
Therefore, using Eqs. (\ref{steven6}) and (\ref{steven8}), the nonvanishing
components of the Ricci tensor are%
\begin{equation}
\mathcal{R}_{11}^{(\text{FR})}=-\frac{1}{2\sigma_{x}^{2}}\text{, and
}\mathcal{R}_{22}^{(\text{FR})}=-\frac{1}{\sigma_{x}^{2}}\text{.}
\label{steven9}%
\end{equation}
Finally, the scalar curvature $\mathcal{R}$ is defined as%
\begin{equation}
\mathcal{R}\overset{\text{def}}{=}\mathcal{R}_{ij}\,g^{ij}=\mathcal{R}%
_{ijk}^{l}\,g_{lm}\,g^{ik}\,g^{jm}\text{.} \label{s14}%
\end{equation}
Therefore, using Eqs. (\ref{steven5}) and (\ref{steven9}), we obtain%
\begin{equation}
\mathcal{R}^{(\text{FR})}=-1\text{.}%
\end{equation}
As a final remark, we recall that the Weyl anisotropy tensor is defined as
\cite{casetti},%
\begin{equation}
W_{ijk}^{l}\overset{\text{def}}{=}\mathcal{R}_{ijk}^{l}-\frac{1}{N-1}\left(
\mathcal{R}_{ik}\delta_{j}^{l}-\mathcal{R}_{ij}\delta_{k}^{l}\right)  \text{,}
\label{weyl}%
\end{equation}
with $W_{lijk}=g^{ll}W_{ijk}^{l}$. Substituting Eqs. (\ref{ciao1}) and
(\ref{steven9}) into Eq. (\ref{weyl}), it happens that the Weyl anisotropy
tensor components are identically zero. Moreover, the sectional curvature is
constant and equals
\begin{equation}
\mathcal{K}^{\left(  \text{FR}\right)  }\overset{\text{def}}{=}\mathcal{R}%
_{1212}/\det\left[  g^{\left(  \text{FR}\right)  }\right]  =-1/2\text{.}
\label{sec1}%
\end{equation}
Therefore, being isotropic and homogeneous, the manifold $\left(
\mathcal{M}\text{, }g^{(\text{FR})}\right)  $ is maximally symmetric. Further
technical details on maximally symmetric manifolds appear in Appendix A.

\subsection{The $\alpha$-order metric}

Consider now a single-variable Gaussian probability density function
$p(x\,|\,\mu_{x}$,$\sigma_{x})$ as defined in Eq. (\ref{steven1}). In what
follows, we study the information geometric properties of such a Gaussian
probability space by using the $\alpha$-order metric tensor in Eq.
(\ref{alpha}). Substituting Eq. (\ref{steven1}) into Eq. (\ref{alpha}), we
have
\begin{equation}
\,\,g_{ij}^{(\alpha)}=\frac{1}{\sigma_{x}^{3}}%
\begin{pmatrix}
\frac{1}{4\sqrt{\pi}} & 0\\
0 & \frac{3}{8\sqrt{\pi}}%
\end{pmatrix}
\text{.} \label{s12}%
\end{equation}
Using the line of reasoning outlined in the previous subsection, we obtain
that the affine connection coefficients are%
\begin{equation}
\Gamma_{11}^{1}=0\text{, }\Gamma_{12}^{1}=\Gamma_{21}^{1}=-\frac{3}%
{2\sigma_{x}}\text{, }\Gamma_{22}^{1}=0\text{, }\,\Gamma_{11}^{2}=\frac
{1}{\sigma_{x}}\text{, }\Gamma_{22}^{2}=-\frac{3}{2\sigma_{x}}\text{ and,
}\Gamma_{12}^{2}=\Gamma_{21}^{2}=0\text{.} \label{steven11}%
\end{equation}
Substituting Eq. (\ref{steven11}) into Eq. (\ref{steven7}), the non-vanishing
Riemann curvature tensor components are%
\begin{equation}
\mathcal{R}_{212}^{1}=-\frac{3}{2\sigma_{x}^{2}}\text{, }\mathcal{R}_{221}%
^{1}=\frac{3}{2\sigma_{x}^{2}}\text{, }\mathcal{R}_{112}^{2}=\frac{1}%
{\sigma_{x}^{2}}\text{, }\mathcal{R}_{121}^{2}=-\frac{1}{\sigma_{x}^{2}%
}\text{.} \label{ciao2}%
\end{equation}
Furthermore, using Eqs. (\ref{steven8}) and (\ref{steven11}), the nonvanishing
components of the Ricci tensor are%
\begin{equation}
\mathcal{R}_{11}^{(\alpha)}=-\frac{1}{\sigma_{x}^{2}}\text{, and }%
\mathcal{R}_{22}^{(\alpha)}=-\frac{3}{2\sigma_{x}^{2}}\text{.} \label{s13}%
\end{equation}
Finally, substituting Eqs. (\ref{s12}) and (\ref{s13}) into Eq. (\ref{s14}),
the scalar curvature $\mathcal{R}^{(\alpha)}$ becomes
\begin{equation}
\mathcal{R}^{(\alpha)}=-8\sqrt{\pi}\sigma_{x}\text{.}%
\end{equation}
As a final remark, we observe that substituting Eqs. (\ref{ciao2}) and
(\ref{s13}) into Eq. (\ref{weyl}), it happens that the Weyl anisotropy tensor
components $W_{lijk}$ are identically zero. However, the sectional curvature
is not constant and equals
\begin{equation}
\mathcal{K}^{\left(  \alpha\right)  }\overset{\text{def}}{=}\mathcal{R}%
_{1212}/\det\left[  g^{\left(  \alpha\right)  }\right]  =-4\sqrt{\pi}%
\sigma_{x}\text{.} \label{sec2}%
\end{equation}
Therefore, being isotropic but not homogeneous, the manifold $\left(
\mathcal{M}\text{, }g^{(\alpha)}\right)  $ is not maximally symmetric.

\section{Entropic motion}

Consider a statistical manifold $\mathcal{M}_{s}$ with a metric $g_{ij}$. The
ED is concerned with the following task \cite{catichaED}: given the initial
and final states, what trajectory is the system expected to follow? The answer
happens to be that the expected trajectory is the geodesic that passes through
the given initial and final states. Moreover, the trajectory follows from a
principle of entropic inference, the MrE algorithm
\cite{caticha12,adom06,adom07,adomphd}. The goal of the MrE method is to
update from a prior distribution $q$\ to a posterior distribution
$P(x)$\ given the information that the posterior lies within a certain family
of distributions $p$. The selected posterior $P(x)$\ is that which maximizes
the logarithm relative entropy $\mathcal{S}[p\left\vert q\right.  ]$,%
\begin{equation}
\mathcal{S}[p\left\vert q\right.  ]\overset{\text{def}}{=}-%
{\displaystyle\int}
dxp\left(  x\right)  \log\frac{p\left(  x\right)  }{q\left(  x\right)
}\text{.}%
\end{equation}
We remark that ED is formally similar to other generally covariant theories:
the dynamics is reversible, the trajectories are geodesics, the system
supplies its own notion of an intrinsic time, the motion can be derived from a
variational principle of the form of Jacobi's action principle rather than the
more familiar principle of Hamilton. Roughly speaking, the canonical
Hamiltonian formulation of ED is an example of a constrained
information-dynamics where the information-constraints play the role of
generators of evolution. For further technical details on the ED framework
used here, we refer to \cite{catichaED}.

A geodesic on a $N$-dimensional manifold $\mathcal{M}_{s}$ represents the
maximum probability path a complex dynamical system explores in its evolution
between initial and final macrostates $\theta_{\text{initial}}$ and
$\theta_{\text{final}}$, respectively. Each point of the geodesic represents a
macrostate parametrized by the macroscopic dynamical variables $\theta
\overset{\text{def}}{=}\left(  \theta^{1}\text{,..., }\theta^{N}\right)  $
defining the macrostate of the system. Each component $\theta^{k}$ with
$k=1$,..., $N$ is a solution of the geodesic equation \cite{catichaED},%
\begin{equation}
\frac{d^{2}\theta^{k}}{d\tau^{2}}+\Gamma_{ij}^{k}\frac{d^{2}\theta^{i}}{d\tau
}\frac{d^{2}\theta^{j}}{d\tau}=0\text{,} \label{geodesic}%
\end{equation}
Furthermore, each macrostate $\theta$ is in a one-to-one correspondence with
the probability distribution $p\left(  x|\theta\right)  $. This is a
distribution of the microstates $x$.

\subsection{The Fisher-Rao information metric}

Substituting Eq. (\ref{steven6}) into Eq. (\ref{geodesic}), the two coupled
nonlinear second-order ordinary differential equations (ODEs) to consider
become%
\begin{equation}
\frac{d^{2}\mu_{x}}{d\tau^{2}}-\frac{2}{\sigma_{x}}\frac{d\mu_{x}}{d\tau}%
\frac{d\sigma_{x}}{d\tau}=0\text{ and, }\frac{d^{2}\sigma_{x}}{d\tau^{2}%
}+\frac{1}{2\sigma_{x}}\left(  \frac{d\mu_{x}}{d\tau}\right)  ^{2}-\frac
{1}{\sigma_{x}}\left(  \frac{d\sigma_{x}}{d\tau}\right)  ^{2}=0\text{.}
\label{A1}%
\end{equation}
Let $\dot{\mu}_{x}\overset{\text{def}}{=}\frac{d\mu_{x}}{d\tau}$ and
$\dot{\sigma}_{x}\overset{\text{def}}{=}\frac{d\sigma_{x}\left(  \tau\right)
}{d\tau}$. Then, the first and the second relations in Eq. (\ref{A1}) become%
\begin{equation}
\ddot{\mu}_{x}-2\frac{\dot{\sigma}_{x}}{\sigma_{x}}\dot{\mu}_{x}=0\text{ and,
}\ddot{\sigma}_{x}+\frac{1}{2\sigma_{x}}\dot{\mu}_{x}^{2}-\frac{\dot{\sigma
}_{x}^{2}}{\sigma_{x}}=0\text{,} \label{a}%
\end{equation}
respectively. From the first relation in Eq. (\ref{a}) we observe that%
\begin{equation}
\frac{\ddot{\mu}_{x}}{\dot{\mu}_{x}}=2\frac{\dot{\sigma}_{x}}{\sigma_{x}%
}\text{.} \label{manu}%
\end{equation}
After some simple algebraic manipulations, we get
\begin{equation}
\dot{\mu}_{x}\left(  \tau\right)  =\mathcal{A}\sigma_{x}^{2}\left(
\tau\right)  \text{,} \label{b}%
\end{equation}
where $\mathcal{A}$ is an arbitrary constant. Substituting Eq. (\ref{b}) in
the second relation in Eq. (\ref{a}), we obtain%
\begin{equation}
\text{ }\sigma_{x}\ddot{\sigma}_{x}-\dot{\sigma}_{x}^{2}+\frac{\mathcal{A}%
^{2}}{2}\sigma_{x}^{4}=0\text{.} \label{c}%
\end{equation}
We note that by integrating Eq. (\ref{c}), we find $\sigma_{x}\left(
\tau\right)  $. Then, using Eq. (\ref{b}), we can obtain an expression for
$\mu_{x}\left(  \tau\right)  $. To simplify the notation, let us set
$\sigma_{x}\left(  \tau\right)  =y\left(  \tau\right)  $ and $a\overset
{\text{def}}{=}\frac{\mathcal{A}^{2}}{2}\in%
\mathbb{R}
_{0}^{+}$. Then, Eq. (\ref{c}) becomes%
\begin{equation}
\text{ }y\ddot{y}-\dot{y}^{2}+ay^{4}=0\text{.} \label{d}%
\end{equation}
To integrate Eq. (\ref{d}), let us consider a first change of variables
\begin{equation}
y\left(  \tau\right)  \overset{\text{def}}{=}\frac{dx\left(  \tau\right)
}{d\tau}=\dot{x}\left(  \tau\right)  \text{.} \label{f}%
\end{equation}
Substituting Eq. (\ref{f}) into Eq. (\ref{d}), we get%
\begin{equation}
\dot{x}\dddot{x}-\ddot{x}^{2}+a\dot{x}^{4}=0\text{.} \label{q}%
\end{equation}
To integrate Eq. (\ref{q}), let us take into consideration a second change of
variables%
\begin{equation}
\dot{x}=\frac{dx\left(  \tau\right)  }{d\tau}\overset{\text{def}}{=}z\left(
x\right)  \text{.} \label{g}%
\end{equation}
Defining $z^{\prime}\overset{\text{def}}{=}dz/dx$, it follows from Eq.
(\ref{g}) that%
\begin{equation}
\ddot{x}=zz^{\prime}\text{ and, }\dddot{x}=\left(  z^{\prime\prime}%
z+z^{\prime2}\right)  z\text{,} \label{h}%
\end{equation}
since,%
\begin{equation}
\ddot{x}=\frac{d\dot{x}}{dt}=\frac{dz}{dt}=\frac{dz}{dx}\frac{dx}{dt}\text{.}%
\end{equation}
Substituting Eqs. (\ref{h}) and (\ref{g}) into Eq.(\ref{q}), we get%
\begin{equation}
z^{\prime\prime}+az=0\text{.} \label{1}%
\end{equation}
A simple integration of Eq. (\ref{1}) yields,%
\begin{equation}
z\left(  x\right)  =c_{1}\sin\left(  \sqrt{a}x+c_{2}\right)  \text{,}
\label{zx}%
\end{equation}
where $c_{1}$ and $c_{2}$ are two \emph{real} integration coefficients.
Recalling that $\dot{x}=z\left(  x\right)  $, we deduce from Eq. (\ref{zx})
that
\begin{equation}
\int^{x}\frac{1}{c_{1}\sin\left(  \sqrt{a}x^{\prime}+c_{2}\right)  }%
dx^{\prime}=\int^{\tau}d\tau^{\prime}\text{.} \label{a11}%
\end{equation}
Observe that,%
\begin{equation}
\frac{d\left\{  \log\left[  \tan\left(  \frac{x}{2}\right)  \right]  \right\}
}{dx}=\frac{1+\tan^{2}\left(  \frac{x}{2}\right)  }{2\tan\left(  \frac{x}%
{2}\right)  }=\frac{1}{\sin\left(  x\right)  }\text{.}%
\end{equation}
Then, upon integration, Eq. (\ref{a11}) yields%
\begin{equation}
\frac{1}{\sqrt{a}c_{1}}\log\left[  \tan\left(  \frac{\sqrt{a}x+c_{2}}%
{2}\right)  \right]  =\tau+c_{3}\text{,} \label{xyz}%
\end{equation}
where the integration coefficient $c_{3}\in%
\mathbb{R}
$. Solving Eq. (\ref{xyz}) for $x=x\left(  \tau\right)  $, we obtain%
\begin{equation}
x\left(  \tau\right)  =\frac{1}{\sqrt{a}}\left\{  2\arctan\left(  \exp\left[
\sqrt{a}c_{1}\left(  \tau+c_{3}\right)  \right]  \right)  -c_{2}\right\}
\text{.}%
\end{equation}
Finally, recalling that $\sigma_{x}\left(  \tau\right)  =\dot{x}\left(
\tau\right)  $, the variance becomes%
\begin{equation}
\sigma_{x}\left(  \tau\right)  =\frac{2c_{1}\exp\left(  c_{1}\sqrt{a}%
\tau+c_{1}c_{3}\sqrt{a}\right)  }{1+\exp\left(  2c_{1}\sqrt{a}\tau+2c_{1}%
c_{3}\sqrt{a}\right)  }\text{.} \label{supp}%
\end{equation}
It is straightforward to verify that indeed $\sigma_{x}\left(  \tau\right)  $
in Eq. (\ref{supp}) satisfies the nonlinear ODE in Eq. (\ref{d}). From Eq.
(\ref{b}), we find that $\mu_{x}\left(  \tau\right)  $ equals%
\begin{equation}
\mu_{x}\left(  \tau\right)  =\sqrt{2a}\int^{\tau}\sigma_{x}^{2}\left(
\tau^{\prime}\right)  d\tau^{\prime}\text{.} \label{supp1}%
\end{equation}
Substituting Eq. (\ref{supp}) into Eq. (\ref{supp1}), the mean $\mu_{x}\left(
\tau\right)  $ becomes%
\begin{equation}
\mu_{x}\left(  \tau\right)  =\frac{c_{4}\left[  1+\exp\left(  2c_{1}\sqrt
{a}\tau+2c_{1}c_{3}\sqrt{a}\right)  \right]  -2\sqrt{2}c_{1}}{1+\exp\left(
2c_{1}\sqrt{a}\tau+2c_{1}c_{3}\sqrt{a}\right)  }\text{,} \label{supp3}%
\end{equation}
where the integration coefficient $c_{4}\in%
\mathbb{R}
$. As a simplifying working hypothesis, we consider geodesic paths with
$c_{3}=0$. Furthermore, we assume that the initial conditions are given by
$\mu_{x}\left(  0\right)  =\mu_{0}$ and $\sigma_{x}\left(  0\right)
=\sigma_{0}$. These initial conditions imply that $c_{1}=\sigma_{0}$ and
$c_{4}=\mu_{0}+\sqrt{2}\sigma_{0}$. Finally, letting $\lambda\overset
{\text{def}}{=}\sqrt{a}=\mathcal{A}/\sqrt{2}\in%
\mathbb{R}
_{+}\backslash\left\{  0\right\}  $, the geodesics in Eqs. (\ref{supp})\ and
(\ref{supp3}) become%
\begin{equation}
\sigma_{x}\left(  \tau\right)  =\frac{2\sigma_{0}\exp\left(  \sigma_{0}%
\lambda\tau\right)  }{1+\exp\left(  2\sigma_{0}\lambda\tau\right)  }\text{,}
\label{sigma}%
\end{equation}
and%
\begin{equation}
\mu_{x}\left(  \tau\right)  =\frac{\left(  \mu_{0}+\sqrt{2}\sigma_{0}\right)
\left[  1+\exp\left(  2\sigma_{0}\lambda\tau\right)  \right]  -2\sqrt{2}%
\sigma_{0}}{1+\exp\left(  2\sigma_{0}\lambda\tau\right)  }\text{,} \label{mu}%
\end{equation}
respectively. We remark that it is straightforward to check that indeed the
expression for $\sigma_{x}\left(  \tau\right)  $ and $\mu_{x}\left(
\tau\right)  $ in Eqs. (\ref{sigma}) and (\ref{mu}), respectively, satisfy the
set of coupled nonlinear ODEs in Eq. (\ref{A1}).

\subsection{The $\alpha$-order metric}

Substituting Eq. (\ref{steven11}) into Eq. (\ref{geodesic}), the two coupled
nonlinear second-order ODEs to consider become%
\begin{equation}
\frac{d^{2}\mu_{x}}{d\tau^{2}}-\frac{3}{\sigma_{x}}\frac{d\mu_{x}}{d\tau}%
\frac{d\sigma_{x}}{d\tau}=0\text{ and, }\frac{d^{2}\sigma_{x}}{d\tau^{2}%
}+\frac{1}{\sigma_{x}}\left(  \frac{d\mu_{x}}{d\tau}\right)  ^{2}-\frac{3}%
{2}\frac{1}{\sigma_{x}}\left(  \frac{d\sigma_{x}}{d\tau}\right)
^{2}=0\text{.} \label{utube}%
\end{equation}
Let $\dot{\mu}_{x}\overset{\text{def}}{=}\frac{d\mu_{x}}{d\tau}$ and
$\dot{\sigma}_{x}\overset{\text{def}}{=}\frac{d\sigma_{x}\left(  \tau\right)
}{d\tau}$. Then, the first and the second relations in Eq. (\ref{utube}) can
be rewritten as%
\begin{equation}
\sigma_{x}\ddot{\mu}_{x}-3\dot{\sigma}_{x}\dot{\mu}_{x}=0\text{ and, }%
\sigma_{x}\ddot{\sigma}_{x}+\dot{\mu}_{x}^{2}-\frac{3}{2}\dot{\sigma}_{x}%
^{2}=0\text{,} \label{u2}%
\end{equation}
respectively. From the first relation in Eq. (\ref{u2}), we note that%
\begin{equation}
\frac{\ddot{\mu}_{x}}{\dot{\mu}_{x}}=3\frac{\dot{\sigma}_{x}}{\sigma_{x}%
}\text{.} \label{manu1}%
\end{equation}
From Eq. (\ref{manu1}), we get
\begin{equation}
\dot{\mu}_{x}\left(  \tau\right)  =\mathcal{A}\sigma_{x}^{3}\left(
\tau\right)  \text{,} \label{u3}%
\end{equation}
where $\mathcal{A}$ is an arbitrary constant. The use of Eq. (\ref{u3}) in the
second relation in Eq. (\ref{u2}) yields%
\begin{equation}
\text{ }\sigma_{x}\ddot{\sigma}_{x}+\mathcal{A}^{2}\sigma_{x}^{6}-\frac{3}%
{2}\dot{\sigma}_{x}^{2}=0\text{.} \label{u4}%
\end{equation}
Observe that by integrating Eq. (\ref{u4}), we find $\sigma_{x}\left(
\tau\right)  $. Then, we can obtain an expression for $\mu_{x}\left(
\tau\right)  $ by employing Eq. (\ref{u3}). Setting $\sigma_{x}\left(
\tau\right)  =y\left(  \tau\right)  $, Eq. (\ref{u4}) becomes%
\begin{equation}
y\ddot{y}+\mathcal{A}^{2}y^{6}-\frac{3}{2}\dot{y}^{2}=0\text{.} \label{u5}%
\end{equation}
To integrate Eq. (\ref{u5}), consider a first change of variables,
\begin{equation}
y\left(  \tau\right)  \overset{\text{def}}{=}\frac{dx\left(  \tau\right)
}{d\tau}=\dot{x}\left(  \tau\right)  \text{.} \label{u6}%
\end{equation}
Substituting Eq. (\ref{u6}) into Eq. (\ref{u5}), we get%
\begin{equation}
\dot{x}\dddot{x}+\mathcal{A}^{2}\dot{x}^{6}-\frac{3}{2}\ddot{x}^{2}=0\text{.}
\label{u7}%
\end{equation}
To integrate Eq. (\ref{u7}), we perform a second change of variables%
\begin{equation}
\dot{x}=\frac{dx\left(  \tau\right)  }{d\tau}\overset{\text{def}}{=}z\left(
x\right)  \text{.} \label{u8}%
\end{equation}
Defining $z^{\prime}\overset{\text{def}}{=}dz/dx$, it follows from Eq.
(\ref{u8}) that%
\begin{equation}
\ddot{x}=zz^{\prime}\text{ and, }\dddot{x}=\left(  z^{\prime\prime}%
z+z^{\prime2}\right)  z\text{.} \label{u9}%
\end{equation}
Substituting Eqs. (\ref{u9}) and (\ref{u8}) into Eq.(\ref{u7}), we get%
\begin{equation}
z^{\prime\prime}z^{3}+\mathcal{A}^{2}z^{6}-\frac{1}{2}z^{2}z^{\prime
2}=0\text{.} \label{u10}%
\end{equation}
To integrate Eq. (\ref{u10}), we propose a third change of variables. Let a
new variable $\omega$ be defined as,%
\begin{equation}
\omega=\omega\left(  z\right)  \overset{\text{def}}{=}z^{\prime}=\frac{dz}%
{dx}\text{.} \label{u11}%
\end{equation}
Using Eq. (\ref{u11}) and noting that $z^{\prime\prime}=\omega^{\prime}\omega$
where, with some abuse of notation, $\omega^{\prime}\overset{\text{def}}%
{=}d\omega/dz$, Eq. (\ref{u10}) becomes an ODE of Bernoulli type \cite{nagle},%
\begin{equation}
\omega^{\prime}-\frac{1}{2z}\omega=-\mathcal{A}^{2}z^{3}\omega^{-1}\text{.}
\label{u12}%
\end{equation}
To integrate Eq. (\ref{u12}), we introduce a fourth change of variables. Let a
new variable $v$ be given by,%
\begin{equation}
v\overset{\text{def}}{=}\omega^{2}\text{.} \label{u13}%
\end{equation}
Then, using Eq. (\ref{u13}) and noting that $2\omega\omega^{\prime}=v^{\prime
}$ with $v^{\prime}\overset{\text{def}}{=}dv/dz$, Eq. (\ref{u12}) becomes%
\begin{equation}
v^{\prime}-\frac{1}{z}v=-2\mathcal{A}^{2}z^{3}\text{.} \label{u14}%
\end{equation}
The most general solution of Eq. (\ref{u14}) is given by,%
\begin{equation}
v\left(  z\right)  =z\left[  -\frac{2}{3}\mathcal{A}^{2}z^{3}+c\right]
\text{,} \label{u15}%
\end{equation}
where $c$ is a \emph{real} integration coefficient. For the sake of
simplicity, we set $c$ equal to zero in what follows and, thus,%
\begin{equation}
v\left(  z\right)  =-\frac{2}{3}\mathcal{A}^{2}z^{4}\text{.} \label{u16a}%
\end{equation}
Using Eqs. (\ref{u16a}), (\ref{u13}), (\ref{u11}), (\ref{u8}), and (\ref{u6})
together with the assumption that $\sigma_{x}\left(  0\right)  =\sigma_{0}$,
we get%
\begin{equation}
\sigma_{x}\left(  \tau\right)  =\frac{\sigma_{0}}{\sqrt{1-\frac{4\mathcal{A}%
_{I}}{6^{1/2}}\sigma_{0}^{2}\tau}}\text{,} \label{u22}%
\end{equation}
where we impose $\mathcal{A}\overset{\text{def}}{=}\left\vert \mathcal{A}%
\right\vert \exp\left(  i\phi_{\mathcal{A}}\right)  $ with $\phi_{\mathcal{A}%
}=-\pi/2$ and $\mathcal{A}_{I}=-\left\vert \mathcal{A}\right\vert <0$. Observe
that $\mathcal{A}=\left\vert \mathcal{A}\right\vert \exp\left(  i\phi
_{\mathcal{A}}\right)  $ with $\phi_{\mathcal{A}}=0$ and $\mathcal{A}%
_{R}=\left\vert \mathcal{A}\right\vert >0$ in the case of the Fisher-Rao
information metric discussed in the previous subsection. At this point, we
note that from a formal mathematical standpoint, $\mu_{x}\left(  \tau\right)
$ is such that $\dot{\mu}_{x}\left(  \tau\right)  =-\Phi\mathcal{A}_{I}%
\sigma_{x}^{3}\left(  \tau\right)  $, where $\Phi$ is a phase factor that
equals $\exp(\left[  -(\pi/2)i\right]  =-i$. Therefore, using Eq. (\ref{u22})
and setting $\mu_{x}\left(  0\right)  =\mu_{0}$, we finally\textbf{
}obtain\textbf{,}%
\begin{equation}
\mu_{x}\left(  \tau\right)  =\mu_{0}+\frac{6^{1/2}\Phi}{2}\sigma_{0}\left(
1-\frac{1}{\sqrt{1-\frac{4\mathcal{A}_{I}}{6^{1/2}}\sigma_{0}^{2}\tau}%
}\right)  \text{.} \label{u23}%
\end{equation}
Observe that $\mu_{x}\left(  \tau\right)  $ in\ Eq. (\ref{u23}) and
$\sigma_{x}\left(  \tau\right)  $ in Eq. (\ref{u22}) satisfy the coupled
system of nonlinear ODEs in Eq. (\ref{utube}). To find a closed form
analytical solution to the \emph{real} geodesic equations when the
distinguishability between two probability distributions is quantified by
means of the $\alpha$-order metric tensor, \emph{complex} geodesic paths were
introduced. Specifically, we employed a nontrivial sequence of suitable change
of variables. In order to reverse these operations and return to the original
variable, we computed a number of indefinite integrals, each one defined up to
a constant of integration.\ Along the way, we have arbitrarily set equal to
zero some of these constants in order to facilitate a return to our starting
point with a closed form expression for the geodesics. However, in so doing,
we were compelled to obtain a \emph{complex} solution for the statistical
variable $\mu_{x}\left(  \tau\right)  $. In what follows, we choose as initial
conditions $\mu_{0}=0$ and $\sigma_{0}=1$ in Eqs. (\ref{sigma}), (\ref{mu}),
(\ref{u22}), and (\ref{u23}). In this case, $\mu_{x}\left(  \tau\right)  $
in\ Eq. (\ref{u23}) becomes purely imaginary.

At this juncture, we emphasize that it is not unusual to employ unphysical
concepts in intermediate steps to obtaining solutions to problems in
theoretical physics \cite{peres04}. For example, to characterize a spacelike
singularity and an event horizon generated by a black hole in the framework of
the AdS/CFT (anti-de-Sitter/conformal field theory) correspondence, it is
convenient to study the boundary-to-boundary correlator expressed in terms of
an expectation value of two operators (two massive fields, for instance)
\cite{veronika1}. In general, when evaluating such a boundary correlator, one
needs to take into consideration multiple geodesics that connect the two
boundary points. In particular, there are scenarios where both real
and\ purely imaginary geodesics can contribute to the computation of the
correlation function \cite{veronika2}. However, despite subtleties related to
the nontrivial mathematical structure of geodesic paths, the boundary
correlator can reveal distinct measurable signals of the black hole singularity.

Motivated by these considerations and desiring the have a computationally
accessible path leading to an analytical closed form solution of the geodesic
trajectories, we allowed for the possibility of considering certain types
of\emph{ complex}\textbf{ }statistical variables as solutions to the geodesic
equation in our information geometric analysis. A couple of remarks are in
order. First, we acknowledge that although the expected value of a\textbf{
}\emph{complex}\textbf{ }random variable of the form $x\overset{\text{def}}%
{=}x_{\mathcal{R}}+ix_{\mathcal{I}}$ involving two\textbf{ }\emph{real}%
\textbf{ }variables $x_{\mathcal{R}}\overset{\text{def}}{=}\operatorname{Re}%
\left(  x\right)  $\textbf{\ }and $x_{\mathcal{I}}\overset{\text{def}}%
{=}\operatorname{Im}\left(  x\right)  $\textbf{\ }is a \emph{complex} number,
in our case $x$\ denotes a \emph{real} random variable. Therefore,
both\textbf{ }$\mu_{x}$\textbf{\ }and $\sigma_{x}$\textbf{\ }assume
\emph{real} values\textbf{. }Second, a rotation in complex analysis is a
one-to-one mapping of the $z$-plane onto the $w$-plane\textbf{ }such that $%
\mathbb{C}
\ni z\mapsto w=ze^{i\varphi}\in%
\mathbb{C}
$\ with $\varphi$\ being a fixed\textbf{ }\emph{real }number. In particular,
we note that the moduli $\left\vert z\left(  \tau\right)  \right\vert $\ and
$\left\vert w\left(  \tau\right)  \right\vert $\ would have the same
asymptotic temporal behavior if they are assumed to be time-dependent
quantities. Therefore, since we were ultimately interested in evaluating the
asymptotic temporal behavior of the information geometric entropy in terms of
the moduli of the statistical variables $\mu_{x}\left(  \tau\right)  $\ and
$\sigma_{x}\left(  \tau\right)  $, we were willing to keep as good solutions
either\textbf{ }\emph{real}\textbf{ }solutions or\textbf{ }\emph{complex}%
\textbf{ }solutions recast\textbf{ }as a\textbf{ }\emph{complex}\textbf{
}constant phase factor times a time-dependent real function to address
otherwise intractable computational issues from an analytical standpoint. For
the sake of scientific honesty, while searching for our geodesic trajectories,
we also give up on any type of generality concerning both initial conditions
and functional form of the statistical macrovariables. Therefore, whenever
needed, we assumed suitable working assumptions that allowed ultimately to
provide a closed form analytical solution for the geodesic equation of the
proposed form in Eqs. (\ref{u22}) and (\ref{u23}). We acknowledge this is
certainly not the most rigorous approach to our problem. We hope to discover a
mathematically rigorous analytical solution to this specific issue in future
efforts. For the time being, we emphasize that the temporal behavior of the
statistical macrovariables obtained in our analytical computations in Eqs.
(\ref{sigma}), (\ref{mu}), (\ref{u22}), and (\ref{u23}) is qualitatively
consistent with the temporal behavior observed after an approximate numerical
integration of the two systems of nonlinear and coupled ODEs in Eqs.
(\ref{A1}) and (\ref{utube}). For the sake of conceptual simplicity, instead
of using the powerful Runge-Kutta method \cite{hildebrand}, we employed a
forward Euler method Matlab code with step size $h=10^{-3}$ in our numerics.
In particular, we numerically verified that%
\begin{equation}
\lim_{\tau\rightarrow\infty}\left[  \frac{\left[  \dot{\mu}_{x}\left(
\tau\right)  \right]  _{\text{FR-metric}}}{\left[  \dot{\mu}_{x}\left(
\tau\right)  \right]  _{\alpha\text{-metric}}}\right]  =0\text{,}%
\end{equation}
that is, $\left[  \mu_{x}\left(  \tau\right)  \right]  _{\text{FR-metric}}$
approaches its limiting constant value at a rate faster than $\left[  \mu
_{x}\left(  \tau\right)  \right]  _{\alpha\text{-metric}}$ approaches its
terminal constant value. Our numerical results for a special choice of initial
conditions are reported in Fig. $1$.

In the next section, we focus on computing the asymptotic temporal behavior of
the information geometric entropy constructed in terms of the moduli of the
statistical macrovariables in Eqs. (\ref{sigma}), (\ref{mu}), (\ref{u22}), and
(\ref{u23}). In this manner, $\mu_{x}\left(  \tau\right)  $ in\ Eq.
(\ref{u23}) and its modulus $\tilde{\mu}_{x}\left(  \tau\right)
\overset{\text{def}}{=}$ $\left\vert \mu_{x}\left(  \tau\right)  \right\vert $
exhibit the same temporal behavior.

\begin{figure}[t]
\centering
\includegraphics[width=0.50\textwidth] {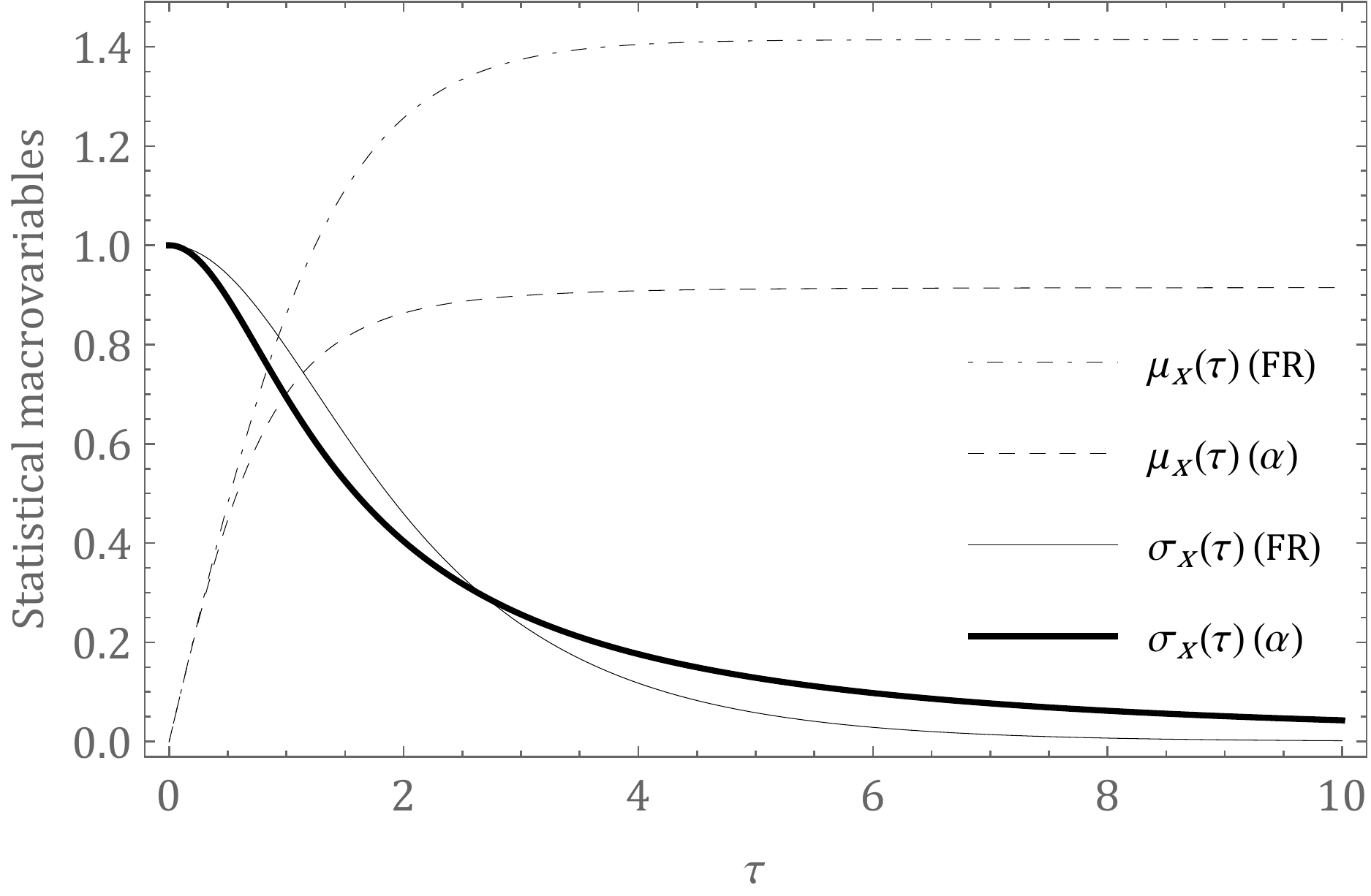}\caption{Plots of numerical
solutions to the geodesic equations for both metrics. A step size of
$h=10^{-3}$ is used, and the initial conditions used in both cases are
$\mu_{x}(0)=0$, $\sigma_{x}(0)=1$, $\dot{\mu}_{x}(0)=1$, and $\dot{\sigma}%
_{x}(0)=0$.}%
\label{fig1}%
\end{figure}

\section{Information Geometric Entropy}

In what follows, we briefly present the concept of the IGE. Assume that the
points $\left\{  p\left(  x\text{; }\theta\right)  \right\}  $ of an
$N$-dimensional curved statistical manifold $\mathcal{M}_{s}$ are parametrized
in terms of $N$ \emph{real} valued variables $\left(  \theta^{1}\text{,...,
}\theta^{N}\right)  $,
\begin{equation}
\mathcal{M}_{s}\overset{\text{def}}{=}\left\{  p\left(  x\text{; }%
\theta\right)  :\theta=\left(  \theta^{1}\text{,..., }\theta^{N}\right)
\in\mathcal{D}_{\boldsymbol{\theta}}^{\left(  \text{tot}\right)  }\right\}
\text{.} \label{smanifold}%
\end{equation}
We remark that the microvariables $x$ in Eq. (\ref{smanifold}) are elements of
the microspace $\mathcal{X}$ while the macrovariables $\theta$ in Eq.
(\ref{smanifold}) belong to the parameter space $\mathcal{D}%
_{\boldsymbol{\theta}}^{\left(  \text{tot}\right)  }$ defined as,
\begin{equation}
\mathcal{D}_{\boldsymbol{\theta}}^{\left(  \text{tot}\right)  }\overset
{\text{def}}{=}{\bigotimes\limits_{j=1}^{N}}\mathcal{I}_{\theta^{j}}=\left(
\mathcal{I}_{\theta^{1}}\otimes\mathcal{I}_{\theta^{2}}\text{...}%
\otimes\mathcal{I}_{\theta^{N}}\right)  \subseteq\mathbb{R}^{N}\text{.}
\label{dtot}%
\end{equation}
The quantity $\mathcal{I}_{\theta^{j}}$ with $1\leq j\leq N$ in Eq.
(\ref{dtot}) is a subset of $\mathbb{R}^{n}$ and specifies the entire range of
permissible values for the statistical macrovariables $\theta^{j}$. The IGE is
a proposed measure of temporal complexity of geodesic paths defined as,
\begin{equation}
\mathcal{S}_{\mathcal{M}_{s}}\left(  \tau\right)  \overset{\text{def}}{=}%
\log\widetilde{vol}\left[  \mathcal{D}_{\boldsymbol{\theta}}\left(
\tau\right)  \right]  \text{,} \label{IGE}%
\end{equation}
where the average dynamical statistical volume\textbf{\ }$\widetilde
{vol}\left[  \mathcal{D}_{\boldsymbol{\theta}}\left(  \tau\right)  \right]  $
is given by,
\begin{equation}
\widetilde{vol}\left[  \mathcal{D}_{\boldsymbol{\theta}}\left(  \tau\right)
\right]  \overset{\text{def}}{=}\frac{1}{\tau}\int_{0}^{\tau}vol\left[
\mathcal{D}_{\boldsymbol{\theta}}\left(  \tau^{\prime}\right)  \right]
d\tau^{\prime}\text{.} \label{rhs}%
\end{equation}
Note that the operation of temporal average is denoted with the tilde symbol
in Eq. (\ref{rhs}). Moreover, the volume\textbf{\ }$vol\left[  \mathcal{D}%
_{\boldsymbol{\theta}}\left(  \tau^{\prime}\right)  \right]  $\textbf{\ }on
the RHS of Eq. (\ref{rhs}) is defined as,
\begin{equation}
vol\left[  \mathcal{D}_{\boldsymbol{\theta}}\left(  \tau^{\prime}\right)
\right]  \overset{\text{def}}{=}\int_{\mathcal{D}_{\boldsymbol{\theta}}\left(
\tau^{\prime}\right)  }\rho\left(  \theta^{1}\text{,..., }\theta^{N}\right)
d^{N}\theta\text{,} \label{v}%
\end{equation}
where $\rho\left(  \theta^{1}\text{,..., }\theta^{N}\right)  $ is the
so-called Fisher density and equals the square root of the determinant
$g\left(  \theta\right)  $ of the metric tensor $g_{ij}\left(  \theta\right)
$,
\begin{equation}
\rho\left(  \theta^{1}\text{,..., }\theta^{N}\right)  \overset{\text{def}}%
{=}\sqrt{g\left(  \theta\right)  }\text{.}%
\end{equation}
We emphasize that the expression of $vol\left[  \mathcal{D}%
_{\boldsymbol{\theta}}\left(  \tau^{\prime}\right)  \right]  $ in Eq.
(\ref{v}) can become more transparent for statistical manifolds with metric
tensor $g_{ij}\left(  \theta\right)  $ whose determinant can be factorized in
the following manner,
\begin{equation}
g\left(  \theta\right)  =g\left(  \theta^{1}\text{,..., }\theta^{N}\right)
={\prod\limits_{k=1}^{N}}g_{k}\left(  \theta^{k}\right)  \text{.}
\label{fattore}%
\end{equation}
With the help of the factorized determinant in Eq. (\ref{fattore}) , the IGE
in Eq. (\ref{IGE}) can be rewritten as
\begin{equation}
\mathcal{S}_{\mathcal{M}_{s}}\left(  \tau\right)  =\log\left\{  \frac{1}{\tau
}\int_{0}^{\tau}\left[  {\prod\limits_{k=1}^{N}}\left(  \int_{\tau_{0}}%
^{\tau_{0}+\tau^{\prime}}\sqrt{g_{k}\left[  \theta^{k}\left(  \xi\right)
\right]  }\frac{d\theta^{k}}{d\xi}d\xi\right)  \right]  d\tau^{\prime
}\right\}  \text{.} \label{IGEmod}%
\end{equation}
We also stress that the leading asymptotic behavior of $\mathcal{S}%
_{\mathcal{M}_{s}}\left(  \tau\right)  $ is used to characterize the
complexity of the statistical models being analyzed. For this reason, it is
customary to take into consideration the quantity
\begin{equation}
\mathcal{S}_{\mathcal{M}_{s}}^{\left(  \text{asymptotic}\right)  }\left(
\tau\right)  \sim\lim_{\tau\rightarrow\infty}\left[  \mathcal{S}%
_{\mathcal{M}_{s}}\left(  \tau\right)  \right]  \text{,}%
\end{equation}
that is to say, the leading asymptotic term in the IGE expression. The
integration space $\mathcal{D}_{\theta}\left(  \tau^{\prime}\right)  $ in Eq.
(\ref{v}) is defined by
\begin{equation}
\mathcal{D}_{\boldsymbol{\theta}}\left(  \tau^{\prime}\right)  \overset
{\text{def}}{=}\left\{  \theta:\theta^{k}\left(  \tau_{0}\right)  \leq
\theta^{k}\leq\theta^{k}\left(  \tau_{0}+\tau^{\prime}\right)  \right\}
\text{,} \label{is}%
\end{equation}
where $\theta^{k}=\theta^{k}\left(  \xi\right)  $ with $\tau_{0}\leq\xi
\leq\tau_{0}+\tau^{\prime}$ and $\tau_{0}$ denoting the initial value of the
affine parameter $\xi$ such that,
\begin{equation}
\frac{d^{2}\theta^{k}\left(  \xi\right)  }{d\xi^{2}}+\Gamma_{ij}^{k}%
\frac{d\theta^{i}}{d\xi}\frac{d\theta^{j}}{d\xi}=0\text{.} \label{ge}%
\end{equation}
The integration domain $\mathcal{D}_{\boldsymbol{\theta}}\left(  \tau^{\prime
}\right)  $ in Eq. (\ref{is}) is an $N$-dimensional subspace of $\mathcal{D}%
_{\boldsymbol{\theta}}^{\left(  \text{tot}\right)  }$ whose elements are
$N$-dimensional macrovariables $\left\{  \theta\right\}  $ with components
$\theta^{k}$ bounded by given limits of integration $\theta^{k}\left(
\tau_{0}\right)  $ and $\theta^{k}\left(  \tau_{0}+\tau^{\prime}\right)  $.
The integration of the $N$-coupled nonlinear second order ODEs in Eq.
(\ref{ge}) determines the temporal functional form of such limits. The IGE at
a certain instant is essentially the logarithm of the volume of the effective
parameter space explored by the system at that instant. The motivation for
considering the temporal average is twofold. In the first case, the temporal
average is used in order to smear out (i.e. average) the possibly highly
complex fine details of the entropic dynamical description of the system on
the manifold. In the second case, the temporal average is used so as to
suppress the consequences of transient effects which may enter the computation
of the expected value of the volume of the effective parameter space. It is
primarily for these two reasons that the the long-term asymptotic temporal
behavior is chosen to serve as an indicator of dynamical complexity. In
summary, the IGE is constructed to furnish an asymptotic coarse-grained
inferential description of the complex dynamics of a system in the presence of
only incomplete information. For further technical details on the IGE, we
refer to Refs. \cite{ali17,felice18,ali18}.

In this section, we wish to compute the asymptotic temporal behavior of
the\textbf{ }information geometric complexity defined as,%
\begin{equation}
\widetilde{IGC}_{\text{asym}}\left(  \tau\right)  \overset{\text{def}}{=}%
\lim_{\tau\rightarrow\infty}\left\{  \frac{1}{\tau}%
{\displaystyle\int\limits_{0}^{\tau}}
\left[
{\displaystyle\int\limits_{\tilde{\mu}_{x}\left(  0\right)  }^{\tilde{\mu}%
_{x}\left(  \tau^{\prime}\right)  }}
{\displaystyle\int\limits_{\tilde{\sigma}_{x}\left(  0\right)  }%
^{\tilde{\sigma}_{x}\left(  \tau^{\prime}\right)  }}
\sqrt{\left\vert g\left(  \tilde{\mu}_{x}\text{, }\tilde{\sigma}_{x}\right)
\right\vert }d\tilde{\mu}_{x}d\tilde{\sigma}_{x}\right]  d\tau^{\prime
}\right\}  \text{,} \label{IGCa}%
\end{equation}
where $IGC_{\text{asym}}\left(  \tau\right)  \overset{\text{def}}{=}%
\exp\left[  \mathcal{S}_{\mathcal{M}_{s}}\left(  \tau\right)  \right]  $,
$\tilde{\mu}_{x}\left(  \tau\right)  \overset{\text{def}}{=}\left\vert \mu
_{x}\left(  \tau\right)  \right\vert \in%
\mathbb{R}
_{+}\backslash\left\{  0\right\}  $ and $\tilde{\sigma}_{x}\left(
\tau\right)  \overset{\text{def}}{=}\left\vert \sigma_{x}\left(  \tau\right)
\right\vert \in%
\mathbb{R}
_{+}\backslash\left\{  0\right\}  $.

\subsection{The Fisher-Rao information metric}

We are interested in the entropic motion from $\tilde{\theta}_{i}%
\overset{\text{def}}{=}\left(  \tilde{\mu}_{x}\left(  \tau_{0}\right)  \text{,
}\tilde{\sigma}_{x}\left(  \tau_{0}\right)  \right)  $ to $\tilde{\theta}%
_{f}\overset{\text{def}}{=}\left(  \tilde{\mu}_{x}\left(  \tau_{\infty
}\right)  \text{, }\tilde{\sigma}_{x}\left(  \tau_{\infty}\right)  \right)  $
with $\tau_{0}=0$ and $\tau_{\infty}=\infty$. Assuming the initial condition
$\left(  \mu_{x}\left(  \tau_{0}\right)  \text{, }\sigma_{x}\left(  \tau
_{0}\right)  \right)  =\left(  \mu_{0}\text{, }\sigma_{0}\right)  =\left(
0\text{, }1\right)  $ and using Eqs. (\ref{sigma}) and (\ref{mu}), we get%
\begin{equation}
\tilde{\mu}_{x}\left(  \tau\right)  =\frac{\sqrt{2}\left(  1+e^{2\lambda\tau
}\right)  -2\sqrt{2}}{1+e^{2\lambda\tau}}\text{, and }\tilde{\sigma}%
_{x}\left(  \tau\right)  =\frac{2e^{\lambda\tau}}{1+e^{2\lambda\tau}}\text{.}
\label{tilda1}%
\end{equation}
The quantity $\lambda$ in\ Eq. (\ref{tilda1})\ is $\lambda\overset{\text{def}%
}{=}\mathcal{A}/\sqrt{2}\in%
\mathbb{R}
_{+}\backslash\left\{  0\right\}  $. Furthermore, using Eq. (\ref{steven3}),
the asymptotic temporal behavior of the information geometric complexity
$\widetilde{IGC}_{\text{asym}}\left(  \tau\right)  $ in\ Eq. (\ref{IGCa})
becomes%
\begin{equation}
\widetilde{IGC}_{\text{asym}}\left(  \tau\right)  \overset{\text{def}}{=}%
\lim_{\tau\rightarrow\infty}\left\{  \frac{1}{\tau}%
{\displaystyle\int\limits_{0}^{\tau}}
\left[
{\displaystyle\int\limits_{\tilde{\mu}_{x}\left(  0\right)  }^{\tilde{\mu}%
_{x}\left(  \tau^{\prime}\right)  }}
{\displaystyle\int\limits_{\tilde{\sigma}_{x}\left(  0\right)  }%
^{\tilde{\sigma}_{x}\left(  \tau^{\prime}\right)  }}
\frac{\sqrt{2}}{\tilde{\sigma}_{x}^{2}}d\tilde{\mu}_{x}d\tilde{\sigma}%
_{x}\right]  d\tau^{\prime}\right\}  \text{.} \label{IGCa1}%
\end{equation}
Using Eq. (\ref{tilda1}), Eq. (\ref{IGCa1}) yields%
\begin{equation}
\widetilde{IGC}_{\text{asym}}\left(  \tau\right)  =\lim_{\tau\rightarrow
\infty}\left\{  \frac{1}{\tau}%
{\displaystyle\int\limits_{0}^{\tau}}
\sqrt{2}\tilde{\mu}_{x}\left(  \tau^{\prime}\right)  \left(  \frac{1}%
{\tilde{\sigma}_{x}\left(  \tau^{\prime}\right)  }-1\right)  d\tau^{\prime
}\right\}  \text{.}%
\end{equation}
After some algebra, we obtain%
\begin{equation}
\widetilde{IGC}_{\text{asym}}\left(  \tau\right)  =\lim_{\tau\rightarrow
\infty}\left\{  \frac{e^{\lambda\tau}}{\lambda\tau}\left[  e^{-2\lambda\tau
}-2e^{-\lambda\tau}\log\left(  1+e^{2\lambda\tau}\right)  +2\lambda\tau
e^{-\lambda\tau}+1\right]  -\frac{2}{\lambda\tau}\left[  1-\log\left(
2\right)  \right]  \right\}  \text{,}%
\end{equation}
that is,
\begin{equation}
\mathcal{S}_{\mathcal{M}_{s}}^{\left(  \text{asymptotic}\right)  }\left(
\tau\right)  =\widetilde{IGE}_{\text{asym}}\left(  \tau\right)  \overset
{\text{def}}{=}\log\left[  \widetilde{IGC}_{\text{asym}}\left(  \tau\right)
\right]  \overset{\tau\rightarrow\infty}{\approx}\tau\text{.} \label{IGE1}%
\end{equation}
Equation (\ref{IGE1}) exhibits asymptotic linear temporal growth of the
information geometric entropy of the statistical model $\left(  \mathcal{M}%
_{s}\text{, }g^{\left(  \text{FR}\right)  }\right)  $.

\subsection{The $\alpha$-order metric}

As previously stated, we are interested in the entropic motion from
$\tilde{\theta}_{i}\overset{\text{def}}{=}\left(  \tilde{\mu}_{x}\left(
\tau_{0}\right)  \text{, }\tilde{\sigma}_{x}\left(  \tau_{0}\right)  \right)
$ to $\tilde{\theta}_{f}\overset{\text{def}}{=}\left(  \tilde{\mu}_{x}\left(
\tau_{\infty}\right)  \text{, }\tilde{\sigma}_{x}\left(  \tau_{\infty}\right)
\right)  $ with $\tau_{0}=0$ and $\tau_{\infty}=\infty$. Assuming the initial
condition $\left(  \mu_{x}\left(  \tau_{0}\right)  \text{, }\sigma_{x}\left(
\tau_{0}\right)  \right)  =\left(  \mu_{0}\text{, }\sigma_{0}\right)  =\left(
0\text{, }1\right)  $ and using Eqs. (\ref{u22}) and (\ref{u23}), we find%
\begin{equation}
\tilde{\mu}_{x}\left(  \tau\right)  =\frac{6^{\frac{1}{2}}}{2}\left(
1-\frac{1}{\sqrt{1-\frac{4\mathcal{A}_{I}}{6^{1/2}}\tau}}\right)  \text{, and
}\tilde{\sigma}_{x}\left(  \tau\right)  =\frac{1}{\sqrt{1-\frac{4\mathcal{A}%
_{I}}{6^{1/2}}\tau}}\text{.} \label{tilda2}%
\end{equation}
The quantity $\mathcal{A}$ in\ Eq. (\ref{tilda2}) is $\mathcal{A}%
\overset{\text{def}}{=}\left\vert \mathcal{A}\right\vert \exp\left(
i\phi_{\mathcal{A}}\right)  $ with $\phi_{\mathcal{A}}=-\pi/2$ and
$\mathcal{A}_{I}=-\left\vert \mathcal{A}\right\vert <0$ (that is,
$\mathcal{A}\overset{\text{def}}{=}i\mathcal{A}_{I}$ with $\mathcal{A}_{I}\in%
\mathbb{R}
_{-}\backslash\left\{  0\right\}  $). Furthermore, using Eq. (\ref{s12}),
$\widetilde{IGC}_{\text{asym}}\left(  \tau\right)  $ in\ Eq. (\ref{IGCa})
becomes%
\begin{equation}
\widetilde{IGC}_{\text{asym}}\left(  \tau\right)  \overset{\text{def}}{=}%
\lim_{\tau\rightarrow\infty}\left\{  \frac{1}{\tau}%
{\displaystyle\int\limits_{0}^{\tau}}
\left[
{\displaystyle\int\limits_{\tilde{\mu}_{x}\left(  0\right)  }^{\tilde{\mu}%
_{x}\left(  \tau^{\prime}\right)  }}
{\displaystyle\int\limits_{\tilde{\sigma}_{x}\left(  0\right)  }%
^{\tilde{\sigma}_{x}\left(  \tau^{\prime}\right)  }}
\sqrt{\frac{3}{32\pi}}\frac{1}{\tilde{\sigma}_{x}^{3}}d\tilde{\mu}_{x}%
d\tilde{\sigma}_{x}\right]  d\tau^{\prime}\right\}  \text{.} \label{IGCa2}%
\end{equation}
Using Eq. (\ref{tilda2}), Eq. (\ref{IGCa2}) yields%
\begin{equation}
\widetilde{IGC}_{\text{asym}}\left(  \tau\right)  =\lim_{\tau\rightarrow
\infty}\left\{  \frac{1}{\tau}%
{\displaystyle\int\limits_{0}^{\tau}}
\frac{1}{2}\sqrt{\frac{3}{32\pi}}\tilde{\mu}_{x}\left(  \tau^{\prime}\right)
\left(  \frac{1}{\tilde{\sigma}_{x}^{2}\left(  \tau^{\prime}\right)
}-1\right)  d\tau^{\prime}\right\}  \text{.}%
\end{equation}
After some algebra, we get%
\begin{equation}
\widetilde{IGC}_{\text{asym}}\left(  \tau\right)  =\lim_{\tau\rightarrow
\infty}\left\{  \frac{1}{8\sqrt{\pi}}\frac{1}{\sqrt{\frac{2^{3/2}}{\sqrt{3}%
}a\tau+1}}+\frac{\sqrt{6}}{16\sqrt{\pi}}a\tau+\frac{\sqrt{6}}{16\sqrt{\pi}%
a}\frac{1}{\sqrt{\frac{2^{3/2}}{\sqrt{3}}a\tau+1}}\frac{1}{\tau}-\frac
{\sqrt{6}}{12\sqrt{\pi}}\frac{a\tau}{\sqrt{\frac{2^{3/2}}{\sqrt{3}}a\tau+1}%
}-\frac{\sqrt{6}}{16\sqrt{\pi}a}\frac{1}{\tau}\right\}  \text{,} \label{z}%
\end{equation}
with $a$ in Eq. (\ref{z}) defined as $a\overset{\text{def}}{=}-\mathcal{A}%
_{I}\in%
\mathbb{R}
_{+}\backslash\left\{  0\right\}  $, that is
\begin{equation}
\widetilde{IGE}_{\text{asym}}\left(  \tau\right)  \overset{\text{def}}{=}%
\log\left[  \widetilde{IGC}_{\text{asym}}\left(  \tau\right)  \right]
\overset{\tau\rightarrow\infty}{\approx}\log\left(  \tau\right)  \text{.}
\label{IGE2}%
\end{equation}
Eq. (\ref{IGE2}) exhibits the asymptotic logarithmic temporal growth of the
information geometric entropy of the statistical model $\left(  \mathcal{M}%
_{s}\text{, }g^{\left(  \alpha\right)  }\right)  $.

\begin{table}[t]
\centering
\begin{tabular}
[c]{c|c|c|c|c}\hline\hline
Metrization & Manifold & IGC growth & IGE growth & Speed of
Convergence\\\hline
Fisher-Rao metric & maximally symmetric & exponential & linear & exponential\\
$\alpha$-order metric & isotropic but nonhomogenous & polynomial &
logarithmic & polynomial\\
&  &  &  &
\end{tabular}
\caption{Asymptotic temporal behavior of the IGC, the IGE, and the speed of
convergence to the final state in the two scenarios being investigated.
Specifically, we consider the entropic motion on a maximally symmetric
(isotropic but nonhomogenous) manifold of Gaussian probability distributions
where distinguishability is quantified by means of the Fisher-Rao information
metric ($\alpha$-metric).}%
\end{table}

\section{Final Remarks}

In this paper, we investigated the effect of distinct metrizations of
probability spaces on the information geometric complexity of entropic motion
on curved statistical manifolds. Specifically, we considered a comparative
analysis based upon Riemannian geometric properties and entropic dynamical
features of a Gaussian probability space where the two dissimilarity measures
between probability distributions were the Fisher-Rao information metric (see
Eq. (\ref{FR})) and the $\alpha$-order entropy metric (see Eq. (\ref{alpha})).
In the former case, we noticed an asymptotic linear temporal growth of the IGE
(see Eq. (\ref{IGE1})) together with a fast convergence to the final state of
the system (see Eq. (\ref{tilda1})). By contrast, in the latter case we
observed an asymptotic logarithmic temporal growth of the IGE (see Eq.
(\ref{IGE2}))\ together with a slow convergence to the final state of the
system (see Eq. (\ref{tilda2})). Our main results are summarized in Fig. 1
together with Table I and can be outlined as follows.

\begin{enumerate}
\item We demonstrated that while $\left(  \mathcal{M}_{s}\text{, }g^{\left(
\text{FR}\right)  }\right)  $ is a maximally symmetric curved statistical
manifold with constant sectional curvature $\mathcal{K}^{\left(
\text{FR}\right)  }$ (see Eq. (\ref{sec1})), the manifold $\left(
\mathcal{M}_{s}\text{, }g^{\left(  \alpha\right)  }\right)  $ is not maximally
symmetric since it is isotropic and nonhomogeneous (see Eq. (\ref{sec2})).

\item We found that the geodesic motion on $\left(  \mathcal{M}_{s}\text{,
}g^{\left(  \text{FR}\right)  }\right)  $ exhibits a fast convergence toward
the final macrostate with $\tilde{\sigma}_{x}\left(  \tau\right)
\overset{\tau\rightarrow\infty}{\propto}\exp\left(  -\lambda\tau\right)  $
with $\lambda\in%
\mathbb{R}
_{+}\backslash\left\{  0\right\}  $ (see Eq. (\ref{tilda1})). Instead, the
geodesic motion on $\left(  \mathcal{M}_{s}\text{, }g^{\left(  \alpha\right)
}\right)  $ shows a slow convergence toward the final macrostate with
$\tilde{\sigma}_{x}\left(  \tau\right)  \overset{\tau\rightarrow\infty
}{\propto}\tau^{-1/2}$ (see Eq.\ (\ref{tilda2})).

\item We determined that the IGE exhibits an asymptotic linear and logarithmic
temporal growth in the case of $\left(  \mathcal{M}_{s}\text{, }g^{\left(
\text{FR}\right)  }\right)  $ and $\left(  \mathcal{M}_{s}\text{, }g^{\left(
\alpha\right)  }\right)  $, respectively. These findings appear in Eqs.
(\ref{IGE1}) and (\ref{IGE2}), respectively.
\end{enumerate}

In addition to having a relevance of its own, our findings can be relevant to
a number of open problems. For instance, thanks to our geodesic motion
analysis together with the observed link between the information geometric
complexity and the speed of convergence to the final state, our work appears
to be useful for deepening our limited understanding about the existence of a
tradeoff between computational speed and availability loss in an information
geometric setting of quantum search algorithms with a thermodynamical flavor
as presented in Refs. \cite{cafaro17,cafaro18}. Furthermore, in view of our
study of the geometrical and dynamical features that emerge from distinct
metrizations of probability spaces, our comparative analysis can help
investigate the unresolved problem of whether the complexity of a convex
combination of two distributions is related to the complexities of the
individual constituents \cite{ay09}. Indeed, unlike the Fisher-Rao information
metric, the $\alpha$-order metric is available in closed form for Gaussian
mixture models \cite{peter06}. We leave the exploration of these intriguing
topics of investigation to future scientific efforts. We also emphasize that
our information geometric analysis shares some resemblance with quantum
cosmological investigations. First, we observe that the connection
coefficients appearing in our information geometric investigation arise from a
symmetric connection (i.e. $T_{ij}^{k}\overset{\text{def}}{=}\Gamma_{ij}%
^{k}-\Gamma_{ji}^{k}=0$, where $T_{ij}^{k}$\ denotes the components of the
torsion tensor \cite{capozziello01}). In principle however, we could
incorporate non-vanishing torsion in our information geometric framework. The
inclusion of torsion could relate in a natural manner to quantum mechanics
given the noncommutative nature of its underlying probabilistic structure. In
particular, given the findings described in our paper, an investigation of the
transition from isotropic to anisotropic features in cosmological models
equipped with torsion \cite{capozziello17} would constitute an intriguing line
of exploration in future information geometric efforts where we quantify the
complexity of statistical models (both isotropic and anisotropic) under
different metrizations. Second, it can be shown in quantum cosmology that a
nonzero cosmological constant can emerge by virtue of using the von Neumann
entropy in cosmological toy models to quantify statistical correlations
between two distinct cosmic epochs (i.e., entanglement between quantum states)
\cite{capozziello13,capozziello11,capozziello13b}. In view of the use of
entropic tools to model, investigate and understand the link between
statistical correlations and quantum entanglement, the aforementioned quantum
cosmological line of research bears a high degree of similarity with our
information geometric complexity characterization of quantum entangled
Gaussian wave packets as presented in Refs. \cite{kim11,kim12}. \ 

Although our considerations are mainly speculative at this time, we hope to
enhance our understanding of the link between quantum cosmological models and
information geometric statistical models in our forthcoming scientific efforts.

\begin{acknowledgments}
C. C. acknowledges the hospitality of the United States Air Force Research
Laboratory (AFRL) in Rome-NY where part of his contribution to this work was
completed. Finally, constructive criticism from an anonymous referee leading
to an improved version of this manuscript are sincerely acknowledged by the authors.
\end{acknowledgments}

\bigskip\pagebreak

\appendix

\section{Maximally symmetric manifolds}

A maximally symmetric manifold must be homogeneous and isotropic
\cite{defelice90,weinberg72}. Homogeneity implies invariance under any
translation along any coordinate axis. Isotropy, instead, implies invariance
under rotation of any coordinate axis into any other coordinate axis. In what
follows, we study the properties of an $N$-dimensional maximally symmetric
manifold in terms of its $N\left(  N+1\right)  /2$ independent Killing
vectors. In particular, we identify the expressions of the scalar curvature
together with the Ricci and Riemann curvature tensors for a maximally
symmetric manifold. It happens that while the homogeneity of the manifold can
be expressed in terms of the behavior of the scalar curvature, the isotropy
feature is encoded in the behavior of the Ricci and Riemann curvature tensors.
In what follows, we use Greek letters to describe the indices of tensorial components.

In terms of the concept of covariant derivative, the Riemann curvature tensor
is defined as%
\begin{equation}
\left[  \nabla_{\gamma}\text{, }\nabla_{\beta}\right]  V_{\alpha}%
\overset{\text{def}}{=}+\mathcal{R}_{\alpha\gamma\beta}^{\delta}V_{\delta
}\text{,} \label{s1}%
\end{equation}
where $V$ denotes an arbitrary vector field, $\left[  \nabla_{\gamma}\text{,
}\nabla_{\beta}\right]  \overset{\text{def}}{=}\nabla_{\gamma}\nabla_{\beta
}-\nabla_{\beta}\nabla_{\gamma}$ is the commutator, and $\mathcal{R}%
_{\alpha\gamma\beta}^{\delta}$ is given by%
\begin{equation}
\mathcal{R}_{\alpha\gamma\beta}^{\delta}\overset{\text{def}}{=}\partial
_{\beta}\Gamma_{\gamma\alpha}^{\delta}-\partial_{\gamma}\Gamma_{\beta\alpha
}^{\delta}+\Gamma_{\beta\lambda}^{\delta}\Gamma_{\gamma\alpha}^{\lambda
}-\Gamma_{\gamma\lambda}^{\delta}\Gamma_{\beta\alpha}^{\lambda}\text{.}
\label{s0}%
\end{equation}
When the vector field is chosen to be a Killing vector $K$, Eq. (\ref{s1})
becomes%
\begin{equation}
\left[  \nabla_{\gamma}\text{, }\nabla_{\beta}\right]  K_{\alpha}%
=\mathcal{R}_{\alpha\gamma\beta}^{\delta}K_{\delta}\text{.}%
\end{equation}
More specifically, however, Killing vectors satisfy the following equation%
\begin{equation}
\nabla_{\alpha}\nabla_{\beta}K_{\gamma}=\mathcal{R}_{\alpha\beta\gamma
}^{\delta}K_{\delta}\text{.} \label{s3}%
\end{equation}
From the associativity of covariant derivatives, we obtain%
\begin{align}
\left[  \nabla_{\gamma}\text{, }\nabla_{\beta}\right]  \nabla_{\delta
}K_{\varepsilon}  &  =\nabla_{\gamma}\nabla_{\beta}\nabla_{\delta
}K_{\varepsilon}-\nabla_{\beta}\nabla_{\gamma}\nabla_{\delta}K_{\varepsilon
}\nonumber\\
& \nonumber\\
&  =\nabla_{\gamma}\left(  \nabla_{\beta}\nabla_{\delta}K_{\varepsilon
}\right)  -\nabla_{\beta}\left(  \nabla_{\gamma}\nabla_{\delta}K_{\varepsilon
}\right)  \text{.} \label{s2}%
\end{align}
Using Eq. (\ref{s3}) and the product rule, Eq. (\ref{s2})\ becomes%
\begin{align}
\left[  \nabla_{\gamma}\text{, }\nabla_{\beta}\right]  \nabla_{\delta
}K_{\varepsilon}  &  =\nabla_{\gamma}\left(  \mathcal{R}_{\beta\delta
\varepsilon}^{\alpha}K_{\alpha}\right)  -\nabla_{\beta}\left(  \mathcal{R}%
_{\gamma\delta\varepsilon}^{\alpha}K_{\alpha}\right) \nonumber\\
& \nonumber\\
&  =\left(  \nabla_{\gamma}\mathcal{R}_{\beta\delta\varepsilon}^{\alpha
}\right)  K_{\alpha}+\mathcal{R}_{\beta\delta\varepsilon}^{\alpha}%
\nabla_{\gamma}K_{\alpha}-\left(  \nabla_{\beta}\mathcal{R}_{\gamma
\delta\varepsilon}^{\alpha}\right)  K_{\alpha}-\mathcal{R}_{\gamma
\delta\varepsilon}^{\alpha}\nabla_{\beta}K_{\alpha}\nonumber\\
& \nonumber\\
&  =\left(  \nabla_{\gamma}\mathcal{R}_{\beta\delta\varepsilon}^{\alpha
}\right)  K_{\alpha}+\mathcal{R}_{\beta\delta\varepsilon}^{\phi}\nabla
_{\gamma}K_{\phi}-\left(  \nabla_{\beta}\mathcal{R}_{\gamma\delta\varepsilon
}^{\alpha}\right)  K_{\alpha}-\mathcal{R}_{\gamma\delta\varepsilon}^{\phi
}\nabla_{\beta}K_{\phi}\text{,}%
\end{align}
that is,%
\begin{equation}
\left[  \nabla_{\gamma}\text{, }\nabla_{\beta}\right]  \nabla_{\delta
}K_{\varepsilon}=\left(  \nabla_{\gamma}\mathcal{R}_{\beta\delta\varepsilon
}^{\alpha}\right)  K_{\alpha}+\mathcal{R}_{\beta\delta\varepsilon}^{\phi
}\nabla_{\gamma}K_{\phi}-\left(  \nabla_{\beta}\mathcal{R}_{\gamma
\delta\varepsilon}^{\alpha}\right)  K_{\alpha}-\mathcal{R}_{\gamma
\delta\varepsilon}^{\phi}\nabla_{\beta}K_{\phi}\text{.} \label{s4}%
\end{equation}
From Eq. (\ref{s2}), we also have%
\begin{align}
\left[  \nabla_{\gamma}\text{, }\nabla_{\beta}\right]  \nabla_{\delta
}K_{\varepsilon}  &  =\nabla_{\gamma}\nabla_{\beta}\nabla_{\delta
}K_{\varepsilon}-\nabla_{\beta}\nabla_{\gamma}\nabla_{\delta}K_{\varepsilon
}\nonumber\\
& \nonumber\\
&  =\nabla_{\gamma}\nabla_{\beta}\left(  \nabla_{\delta}K_{\varepsilon
}\right)  -\nabla_{\beta}\nabla_{\gamma}\left(  \nabla_{\delta}K_{\varepsilon
}\right)  \text{,}%
\end{align}
that is, using the product rule,%
\begin{align}
\left[  \nabla_{\gamma}\text{, }\nabla_{\beta}\right]  \nabla_{\delta
}K_{\varepsilon}  &  =\left(  \nabla_{\gamma}\nabla_{\beta}\right)
(\nabla_{\delta})K_{\varepsilon}+\left(  \nabla_{\gamma}\nabla_{\beta}\right)
\left(  K_{\varepsilon}\right)  \nabla_{\delta}-\left(  \nabla_{\beta}%
\nabla_{\gamma}\right)  (\nabla_{\delta})K_{\varepsilon}-\left(  \nabla
_{\beta}\nabla_{\gamma}\right)  (K_{\varepsilon})\nabla_{\delta}\nonumber\\
& \nonumber\\
&  =\left(  \nabla_{\gamma}\nabla_{\beta}\right)  (\nabla_{\delta
})K_{\varepsilon}-\left(  \nabla_{\beta}\nabla_{\gamma}\right)  (\nabla
_{\delta})K_{\varepsilon}+\left(  \nabla_{\gamma}\nabla_{\beta}\right)
\left(  K_{\varepsilon}\right)  \nabla_{\delta}-\left(  \nabla_{\beta}%
\nabla_{\gamma}\right)  (K_{\varepsilon})\nabla_{\delta}\nonumber\\
& \nonumber\\
&  =\left[  \nabla_{\gamma}\text{, }\nabla_{\beta}\right]  (\nabla_{\delta
})K_{\varepsilon}+\left[  \nabla_{\gamma}\text{, }\nabla_{\beta}\right]
(K_{\varepsilon})\nabla_{\delta}\text{,}%
\end{align}
so that%
\begin{equation}
\left[  \nabla_{\gamma}\text{, }\nabla_{\beta}\right]  \nabla_{\delta
}K_{\varepsilon}=\left[  \nabla_{\gamma}\text{, }\nabla_{\beta}\right]
(\nabla_{\delta})K_{\varepsilon}+\left[  \nabla_{\gamma}\text{, }\nabla
_{\beta}\right]  (K_{\varepsilon})\nabla_{\delta}\text{.} \label{s5}%
\end{equation}
Using Eq. (\ref{s1}), Eq. (\ref{s5}) becomes%
\begin{equation}
\left[  \nabla_{\gamma}\text{, }\nabla_{\beta}\right]  \nabla_{\delta
}K_{\varepsilon}=\mathcal{R}_{\delta\beta\gamma}^{\alpha}\nabla_{\alpha
}K_{\varepsilon}+\mathcal{R}_{\varepsilon\beta\gamma}^{\alpha}K_{\alpha}%
\nabla_{\delta}\text{.} \label{s6}%
\end{equation}
Recalling that the commutator of two constant vectors is zero, Eq. (\ref{s6})
yields%
\begin{equation}
\left[  \nabla_{\gamma}\text{, }\nabla_{\beta}\right]  \nabla_{\delta
}K_{\varepsilon}=\mathcal{R}_{\delta\beta\gamma}^{\alpha}\nabla_{\alpha
}K_{\varepsilon}+\mathcal{R}_{\varepsilon\beta\gamma}^{\alpha}\nabla_{\delta
}K_{\alpha}\text{.} \label{s7}%
\end{equation}
Equating Eq. (\ref{s4}) and Eq\textbf{.} (\ref{s7}), we obtain%
\begin{equation}
\left(  \nabla_{\gamma}\mathcal{R}_{\beta\delta\varepsilon}^{\alpha}\right)
K_{\alpha}+\mathcal{R}_{\beta\delta\varepsilon}^{\phi}\nabla_{\gamma}K_{\phi
}-\left(  \nabla_{\beta}\mathcal{R}_{\gamma\delta\varepsilon}^{\alpha}\right)
K_{\alpha}-\mathcal{R}_{\gamma\delta\varepsilon}^{\phi}\nabla_{\beta}K_{\phi
}=\mathcal{R}_{\delta\beta\gamma}^{\alpha}\nabla_{\alpha}K_{\varepsilon
}+\mathcal{R}_{\varepsilon\beta\gamma}^{\alpha}\nabla_{\delta}K_{\alpha
}\text{,}%
\end{equation}
that is,%
\begin{equation}
\left(  \nabla_{\gamma}\mathcal{R}_{\beta\delta\varepsilon}^{\alpha}%
-\nabla_{\beta}\mathcal{R}_{\gamma\delta\varepsilon}^{\alpha}\right)
K_{\alpha}+\mathcal{R}_{\beta\delta\varepsilon}^{\phi}\nabla_{\gamma}K_{\phi
}-\mathcal{R}_{\gamma\delta\varepsilon}^{\phi}\nabla_{\beta}K_{\phi
}-\mathcal{R}_{\varepsilon\beta\gamma}^{\alpha}\nabla_{\delta}K_{\alpha
}-\mathcal{R}_{\delta\beta\gamma}^{\alpha}\nabla_{\alpha}K_{\varepsilon
}=0\text{.} \label{s8}%
\end{equation}
Since Killing vectors satisfy the relation $\nabla_{\alpha}K_{\varepsilon
}+\nabla_{\varepsilon}K_{\alpha}=0$, introducing the Kronecker delta, Eq.
(\ref{s8}) can be recast as%
\begin{equation}
\left(  \nabla_{\gamma}\mathcal{R}_{\beta\delta\varepsilon}^{\alpha}%
-\nabla_{\beta}\mathcal{R}_{\gamma\delta\varepsilon}^{\alpha}\right)
K_{\alpha}+\left[  \mathcal{R}_{\beta\delta\varepsilon}^{\phi}\delta_{\gamma
}^{\alpha}-\mathcal{R}_{\gamma\delta\varepsilon}^{\phi}\delta_{\beta}^{\alpha
}+\mathcal{R}_{\varepsilon\beta\gamma}^{\alpha}\delta_{\delta}^{\phi
}-\mathcal{R}_{\delta\beta\gamma}^{\alpha}\delta_{\varepsilon}^{\phi}\right]
\nabla_{\alpha}K_{\phi}=0\text{.} \label{s9}%
\end{equation}
In general, the quantities $K_{\alpha}$ and $\nabla_{\alpha}K_{\phi}$ cannot
be prescribed independently. However, when a manifold is maximally symmetric
and admits all the allowed Killing forms, the quantities $K_{\alpha}$ and
$\nabla_{\alpha}K_{\phi}$ can be prescribed independently. In order for the
sets $\left\{  K_{\alpha}\right\}  $ and $\left\{  \nabla_{\alpha}K_{\phi
}\right\}  $ to be independently specifiable, we impose%
\begin{equation}
\nabla_{\gamma}\mathcal{R}_{\beta\delta\varepsilon}^{\alpha}-\nabla_{\beta
}\mathcal{R}_{\gamma\delta\varepsilon}^{\alpha}=0 \label{s10}%
\end{equation}
and,%
\begin{equation}
\mathcal{R}_{\beta\delta\varepsilon}^{\phi}\delta_{\gamma}^{\alpha
}-\mathcal{R}_{\gamma\delta\varepsilon}^{\phi}\delta_{\beta}^{\alpha
}+\mathcal{R}_{\varepsilon\beta\gamma}^{\alpha}\delta_{\delta}^{\phi
}-\mathcal{R}_{\delta\beta\gamma}^{\alpha}\delta_{\varepsilon}^{\phi
}=0\text{.} \label{s11}%
\end{equation}
For an $N$-dimensional curved manifold, a suitable sequence of tensor algebra
manipulations of Eq. (\ref{s11}) leads to the following expressions of the
Ricci and the Riemann curvature tensors,%
\begin{equation}
\mathcal{R}_{\alpha\delta}=\frac{1}{N}g_{\alpha\delta}\mathcal{R}\text{, and
}\mathcal{R}_{\beta\alpha\varepsilon\delta}=\frac{\mathcal{R}}{N\left(
N-1\right)  }\left(  g_{\alpha\varepsilon}g_{\delta\beta}-g_{\alpha\delta
}g_{\varepsilon\beta}\right)  \text{,} \label{s12a}%
\end{equation}
respectively. In Eq. (\ref{s12a}), $g_{\alpha\delta}$ and $\mathcal{R}$ denote
the metric tensor and the scalar curvature of the manifold, respectively.
Finally, using the second relation in Eq. (\ref{s12a}) together with a
convenient sequence of tensor algebra manipulations, Eq. (\ref{s10}) yields%
\begin{equation}
\nabla_{\phi}\mathcal{R}=0\text{.} \label{s13a}%
\end{equation}
Eq. (\ref{s13a}) implies that the scalar curvature must be covariantly
constant for a maximally symmetric manifold. The relations in Eq. (\ref{s12a})
are valid for an isotropic manifold while Eq. (\ref{s13a}) holds true for a
homogeneous manifold. For a maximally symmetric manifold, both Eqs.
(\ref{s12a}) and (\ref{s13a}) must hold true.

We point out that instead of using Eq. (\ref{s1}), it is possible to define
the Riemann curvature tensor by the relation
\begin{equation}
\left[  \nabla_{\gamma}\text{, }\nabla_{\beta}\right]  V_{\alpha}%
\overset{\text{def}}{=}-\mathcal{R}_{\alpha\gamma\beta}^{\delta}V_{\delta
}\text{.} \label{s14a}%
\end{equation}
In this case, the Riemannian curvature tensor components have the opposite
sign compared to those defined in Eq. (\ref{s0}). In particular, in this case
the second relation in Eq. (\ref{s12a}) becomes%
\begin{equation}
\mathcal{R}_{\beta\alpha\varepsilon\delta}=\frac{\mathcal{R}}{N\left(
N-1\right)  }\left(  g_{\alpha\delta}g_{\varepsilon\beta}-g_{\alpha
\varepsilon}g_{\delta\beta}\right)  \text{.} \label{s15}%
\end{equation}
As a final remark, we recall that the Weyl anisotropy tensor is defined as
\cite{casetti},%
\begin{equation}
W_{\alpha\varepsilon\delta}^{\beta}\overset{\text{def}}{=}\mathcal{R}%
_{\alpha\varepsilon\delta}^{\beta}-\frac{1}{N-1}\left(  \mathcal{R}%
_{\alpha\delta}\delta_{\varepsilon}^{\beta}-\mathcal{R}_{\alpha\varepsilon
}\delta_{\delta}^{\beta}\right)  \text{.}%
\end{equation}
In the working assumption of an isotropic manifold, using the first relation
in Eq. (\ref{s12a}) and contracting $W_{\alpha\varepsilon\delta}^{\beta}$ with
$g_{\beta\beta}$, we\textbf{ }obtain%
\begin{equation}
g_{\beta\beta}W_{\alpha\varepsilon\delta}^{\beta}=g_{\beta\beta}%
\mathcal{R}_{\alpha\varepsilon\delta}^{\beta}-\frac{\mathcal{R}}{N\left(
N-1\right)  }\left(  g_{\alpha\delta}g_{\beta\beta}\delta_{\varepsilon}%
^{\beta}-g_{\alpha\varepsilon}g_{\beta\beta}\delta_{\delta}^{\beta}\right)
\text{,}%
\end{equation}
that is, by means of Eq.\ (\ref{s15}),
\begin{equation}
W_{\beta\alpha\varepsilon\delta}=\mathcal{R}_{\beta\alpha\varepsilon\delta
}-\frac{\mathcal{R}}{N\left(  N-1\right)  }\left(  g_{\alpha\delta}%
g_{\beta\varepsilon}-g_{\alpha\varepsilon}g_{\delta\beta}\right)  =0
\label{s16}%
\end{equation}
From Eq. (\ref{s16}), we conclude that the Weyl anisotropy tensor vanishes in
the case of an isotropic manifold.

\end{document}